# Analysing contact conditions in hard-on-hard hip replacements: effectiveness of current analytical methods and novel data-driven approach

K. Nitish Prasad, M. Abhilash, P. Ramkumar*

Advanced Tribology Research Lab (ATRL), Machine Design Section, Department of Mechanical Engineering,

Indian Institute of Technology Madras (IITM), Chennai, India

*Corresponding author: ramkumar@iitm.ac.in

**Abstract**

Contact mechanics models that can accurately estimate the contact conditions and, consequently predicting wear must be developed for hip implants. This study analyses and verifies the existing Hertz and Fang analytical models applicable to hard-on-hard hip implants. The contact parameters, such as the maximum contact pressure, contact radius and the maximum deformation, are considered for the validation with FEM. Both analytical models fail to predict the contact conditions throughout a gait cycle. A novel data-driven ANN model is developed to comprehensively predict the contact conditions considering different input parameters. The analysis show that cup thickness significantly affects the output contact conditions. Therefore, it is recommended to consider cup thickness in the analytical model interpretation for hard-on-hard hip implants.

**Nomenclature**

| | |
|---|---|
| $A$ | Projected contact half-radius |
| $R_1$ | Radius of the femoral head |
| $D_1$ | Diameter of the femoral head |
| $R_2$ | Radius of the acetabular cup |
| $W$ | Applied load |
| $\Delta R$ | Radial clearance |
| $\delta$ | Maximum elastic deformation |
| $v_1$ | Poisson ratio of the femoral head |
| $v_2$ | Poisson ratio of the acetabular cup |
| $E_1$ | Young's modulus of the femoral head |
| $E_2$ | Young's modulus of the acetabular cup |
| N | Dimensionless exponential coefficient |
| r | Projected contact half-radius from the centre throughout the contact profile along the symmetry axis ($0 \leq r \leq a$) |
| $P_m$ | Maximum contact pressure |
| $P(r)$ | Pressure distribution throughout the contact profile |
| B, C, g | Correlation coefficients |
| $\Gamma$ | Gamma function |
| T | Thickness of the acetabular cup |

## 1. Introduction

Ball-on-socket joints have huge applications in mechanical design particularly in automotive and biomedical fields. Among the biomedical applications, the focus is given on the artificial hip implant in this study which is a ball-on-socket joint having three degrees of freedom aiding human mobility [1–3]. During a total hip replacement (THR) procedure, the hip implant is used to replace the natural hip joint. The hip implant uses body fluid as a lubricant to keep the synovial capsule intact. The acetabular cup and femoral head of the implant act as the contact tribo-pair. Among the two different material configurations (hard-on-soft and hard-on-hard), hard-on-hard tribo-pair combinations such as Metal-on-Metal (MoM) and Ceramic-on-Ceramic (CoC) are selected in this study as it offers less wear for improved longevity [4–6]. Despite their global widespread use, hard-on-hard hip implants fail prematurely in vivo as a result of aseptic loosening caused due to friction and wear [7–9]. Therefore, analysing the contact conditions is important in hip implants as it helps to predict the wear and operating lubricating regime of the implants. Like all the other mechanical components, the contact tribo-pair in hard-on-hard hip implants have a suitable clearance between the components to facilitate freedom of movement with the interaction [10]. This arrangement makes the contact to be nearly conformal between the tribo-pair. Further, the contact area is also estimated to be a circular profile when the contact occurs inside the bearing surface of the cup and no major offset between the cup and head centres, that is, no microseparation [11–13].

Contact mechanics modelling is one of the key areas, where the application of analytical models needs to be developed as current solution providers take enormous computational time due to the nonlinearity [14]. The primary output from the contact models, such as contact pressure, contact profile and maximum deformation, is helpful in estimating the mechanical failure modes, particularly wear and fatigue, which will avoid premature failures in the interacting components [15,16]. The models must be able to estimate the contact pressure profile as it will help for comprehensive contact and wear study [17]. In addition, the maximum deformation from the contact tribo-pair will help in determining the magnitude of EHL regime modelling [18,19]. To improve wear prediction, a thorough contact analysis is required in order to pinpoint the precise location and amount of wear that is occurring between the two surfaces [1]. Until now, researchers have widely used direct numerical simulations to predict and analyse the comprehensive contact and wear analysis. However, no analytical models are developed specifically for hard-on-hard hip implants to directly predict the contact phenomena in nearly conformal contacts.

At present, some of the existing theoretical models are used to evaluate the contact pressure thereby predicting the wear in ball-on-socket joints. In these, a very few models are developed particularly for the nearly conformal and fully conformal contacts. Hertz developed a closed form analytical solution for non-conformal contact sphere on flat surfaces. The developed model had a frictionless contact assumption with small deformations. Until now, the Hertz model has been widely used for analytical prediction of contact pressure and specifically used for mesh convergence studies alone in a wide range of literature studies [20–23] in ball-on-socket joints including hip implants. As the ball-on-socket joint becomes nearly conformal under minimum clearance conditions, the validity of this model was tested and subsequently modified in further studies. Zhangang Sun and Caizhe Hao [24] performed the validation of Hertz model to ball-on-socket joints with a fixed load, fixed radius of curvature and different ball diameters using 2D FEM axisymmetric models. It was found that when the ratio of curvature radius to ball diameter less than 0.536, the Hertz model becomes invalid. Steuermann [15] proposed a model based on 2D contacts and established the pressure distribution to be an even order polynomial function. This model improved the Hertz accuracy in pressure distribution, but it failed at lower clearances as it still follows elastic half-space approximation. Fang et.al [25] proposed and developed a universal approximate semi-analytical model for both conformal and non-conformal contact of spherical surfaces. The proposed model is not only limited to an elastic half-space but also can be universally used to calculate the pressure distribution of conformal and non-conformal contacts. There is a basic need to comprehend how well the present theoretical models can predict contact conditions because the capability to predict contact conditions using these models has not been applied to hip implants.

As the prediction of contact conditions in conformal contacts using analytical models is not accurate, consequently the Finite Element Method (FEM) is used predominantly as it doesn't need any geometrical assumptions and providing accurate results but with the expense of computational cost. The results from the FEM are used to verify and developed new contact theories as described in literature [15,25]. Further, the FEM is widely used over the last decade to analyse the contact profile accurately in hip implants for different geometrical, operational and material configurations [13,26–28]. Thus, in this study the results from the FEM are used to validate and compare the contact results with the existing analytical models.

In the recent past, numerous Machine Learning (ML) models have been developed to analyse the relationships between the variables which helps in the prediction of friction and wear performance of tribo-systems. Specifically, Artificial Neural Network (ANN) models [29–31] have significant contribution in developing the interrelationship and the effect of the input system parameters during the design and operating phase of mechanical systems. Also, several studies have trained ANN model using numerical simulations to predict complex contact conditions with less computational cost. Studies [32,33] have generated load displacement curves and mechanical properties using ANN based on the 2D and 3D FEM simulations of nano indentation tests. Haibo Xie et al. [34] used scratch tests results from FEM and experiments to train an ANN model which can predict friction coefficients. The results reveal the relationships between the scratch process parameters affecting material deformation which aren't determined by numerical simulations. Alper Polat [35] used deep learning neural network to determine contact lengths in a homogeneous elastic layer pressed with two elastic punches. The results were verified with the FEM and the margin of error was less than the threshold limit. Therefore, with the understanding of the significance of ML models, an ANN model will be developed and trained with the FEM results in this study. The developed model will predict the output contact conditions with high accuracy and less computational cost. It can also determine the effect of individual input parameters affecting the output contact variables.

In the present study, the existing popular analytical models i.e. Hertz and Fang which are applicable for ball-on-socket joints are applied to the hard-on-hard hip implants and their capability to predict the contact conditions are verified. The output contact parameters such as contact pressure, contact radius and maximum deformation from these models are validated with the FEM. Further, a data-driven neural network ANN model is trained with the FEM data with predicts the output contact parameters with higher accuracy and less computational cost.

The determination of the pressure distribution index in the contact profile is also investigated in this study. Finally, the relationships obtained for the input and output contact variables will help in developing a new analytical contact mechanics model suitable for hard-on-hard hip implants.

## 2. Model development and Computational Methodology

Under standard loading conditions with a suitable clearance, the femoral head of a hip implant and the acetabular cup make an axisymmetric nearly conformal contact [1,11,13]. The boundary profile of the contact looks like a circle [36]. Though, there is a circular profile contact between the cup-head tribo-pair under standard loading, the size of the contact profile varies upon the magnitude of the load, radial clearance, cup thickness, femoral head size and tribo-pair materials [37,38]. The size of the circular contact profile with the comparison to the radii of curvature of the two surfaces decides the assumption of the type of contact. The following sections will discuss the existing analytical model (Hertz) and semi-analytical model (Fang) as well as the computational approaches of FEM along with neural networks that are applied to the THR joints.

### 2.1 Hertz analytical formulation

Earlier studies [21,39] have applied the well-known, traditional Hertz theory of contact to the calculations for wear in hip implants. The Hertz's theory was formulated taking into account the elastic contact ball-on-plane configuration, which is non-conformal when using the half-space approximation. Studies [22,39,40] further state that until the head comes in contact with the cup at a higher inclination angle (>55°), the assumption of a ball-on-plane structure is valid for hard-on-hard material tribo-pairs in hip implants. This is because there will be less area of contact in relation to the radii of curvature of the two contact surfaces. The contact radius, maximum contact pressure, contact pressure distribution, and maximum elastic deformation between the two bodies are expressed in Eqs.(1)-(4). Hertz also assumed that the pressure profile variation as a parabola.

$$a = \sqrt[3]{\frac{3WR_{eq}}{4E_{eq}}} \tag{1}$$

$$P_m = \frac{3W}{2\pi a^2} \tag{2}$$

$$P(r) = P_m \left(1 - \frac{r^2}{a^2}\right)^{\frac{1}{2}} \tag{3}$$

$$\delta = \frac{a^2}{R_{eq}} \tag{4}$$

The equivalent radius ($R_{eq}$) and equivalent modulus ($E_{eq}$) are defined as shown in Eq. (5) and (6) respectively.

$$R_{eq} = \frac{R_1 R_2}{\Delta R} \tag{5}$$

$$E_{eq} = \left[\frac{1 - v_1^2}{E_1} + \frac{1 - v_2^2}{E_2}\right]^{-1} \tag{6}$$

However, the size of the contact profile depends on various parameters such as radial clearance, thickness etc. as discussed earlier. If the radial clearance between the two surfaces gets sufficiently small ($\Delta R \to 0$), size of the contact profile increases, and the contact radius starts approaching towards the radius of the contacting head. Therefore, consideration of the half-space approximation will no longer be valid, and the Hertz model might not be suitable for the above case.

**2.2 Fang semi-analytical formulation**

Considering the limitations of the Hertz formulation, Fang et al. [25] proposed an semi-analytical model suitable for conformal as well as non-conformal contacts in elastic spherical cavity contacts. The proposed model overcomes the limitations of Hertz by ignoring the elastic half-space assumption. The contact pressure distribution equation Eq. (12) has an exponent $n$ and the value for that exponent was determined with the results of FEM for steel-on-steel contacts. In case of highly non-conformal contacts the exponent $n$ becomes 0.5, which states that the pressure distribution equation becomes similar to Hertz model as shown in Eq. (3). It is important to remember that the contact radius will become infinity when $\Delta R \to 0$ in the Hertz model. The introduction of coefficient functions $C$, $B$ and $g$ in the Fang model prevents this condition and causes the contact radius approach towards $R_1$ which is realistic. Further, the proposed Fang model [25] had the results verified with FEM for the steel-steel and steel-beryllium bronze contacts. The model was tested with high loads (80 – 1900) kN and ball radii (20, 50 and 100 mm). Additionally, a new parameter named as 'contact radius proportion' (CRP) and it was determined along with the proposed model, which is the ratio of the projected radius of contact profile *a* to the socket radius $R_2$. The authors also proposed that Hertz model will be valid only if CRP is less than 0.2, beyond which the relative error will significantly increase. The contact radius, maximum contact pressure, contact pressure distribution, and maximum elastic deformation along with the coefficient functions between the two bodies in the Fang model are expressed in Eqs. (7)-(14).

$$a = \left[\frac{4BR_1R_2(k_1 + k_2)W}{\pi(g\Delta R + C)}(n + 1/2)(n + 1)\right]^{\frac{1}{3}} \tag{7}$$

$$n = 0.5 - 0.24 \exp\left[-15.08\left(1 - \frac{a}{R_2}\right)\right] \tag{8}$$

$$B = \frac{\sqrt{\pi}\Gamma(n + 1)}{2\Gamma(3/2 + n)} \tag{9}$$

$$C = \frac{3.8304B(k_1 + k_2)W}{\pi R_2} \tag{10}$$

$$g(a) = \frac{2}{\pi} + \left(\frac{a}{R_2}\right)^2 \tag{11}$$

$$P(r) = P_m\left(1 - \frac{r^2}{a^2}\right)^n \tag{12}$$

$$P_m = (n + 1)\frac{W}{\pi a^2} \tag{13}$$

$$\delta = 4(k_1 + k_2)BP_m a \tag{14}$$

Where $k_1 = \frac{1-v_1^2}{\pi E_1}$ and $k_2 = \frac{1-v_2^2}{\pi E_2}$

However, as the present study is limited to the application to hip replacements, the model will be tested at lower load magnitudes and geometrical parameters similar to the clinical sizes.

## 2.3 Material, Geometrical and Loading parameters

Based on the objectives of the present study, two different material combinations are considered, i.e., MoM and CoC. The materials and their properties considering each configuration are tabulated in the Table 1.

**Table 1**

Input material properties

| Tribo-pair | Material | Density (g/cm$^3$) | Young's Modulus $E$ (GPa) | Poisson's ratio $v$ |
|---|---|---|---|---|
| MoM | CoCrMo | 8.30 | 210 | 0.30 |
| CoC | ZTA (Biolox Delta) | 4.37 | 350 | 0.26 |

The geometrical parameters considered for this study are taken from the literature [22] which describes the actual geometry of the hip implants in clinical use and are shown in Table 2. Three different head diameters (28, 32, and 36 mm) are selected to accommodate a variety of patient requirements and implant suppliers. [20]. Four different thicknesses (from 5 to 11 mm) of the cup and six different radial clearances (from 10 to 150 µm) are considered for the analysis. The same geometrical parameters are taken into account for both material tribo-pairs because the study is focused on the investigation and applicability of the existing analytical models and a comparison investigation between the two material configurations.

**Table 2**

Geometrical parameters for the analysis

| Thickness of the cup, $t$ (mm) | 5, 7, 9 and 11 |
|---|---|
| Femoral head diameter, $D_1$ (mm) | 28, 32 and 36 |
| Radial clearance, $\Delta R$ (µm) | 10, 30, 50, 75, 100 and 150 |

The loads and angular displacements are picked from ISO 14242-1, which is considered as a popular reference standard and resembles the standard normal walking cycle as shown in Fig. 1 [22,41]. The load cycle follows a twin-peak pattern with a load range between 300 N and 3000 N. The minimum load acts during the swing phase of the cycle. The first 60% duration of the total load cycle comprises of the stance phase followed by the remaining swing phase. The input ISO 14242-1 gait cycle is split into 40 steps to get smooth contact pressure profile generation based on the slide track convergence test obtained from published studies [1,13].

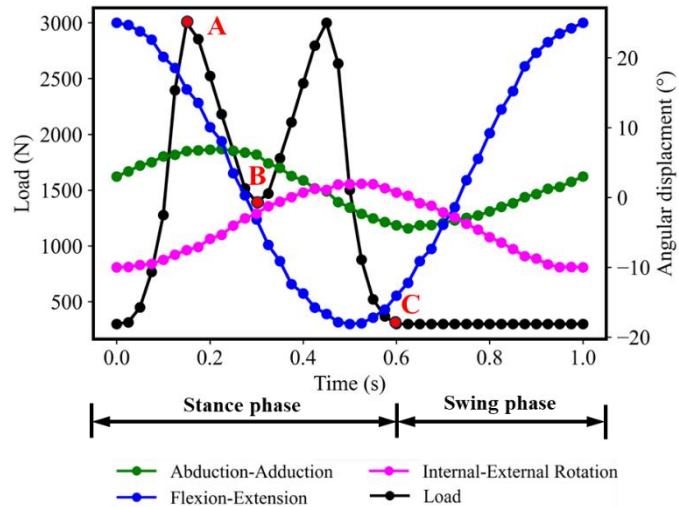

**Fig. 1.** ISO 14242-1 Loads and Rotations

## 2.4 FEM Analysis

As this study involves the applicability of the existing analytical models to THR, the contact parameters results (pressure, radius, and maximum deformation) are verified with FEM. The analyses are performed on a typical ball-on-socket model similar to the head and cup in THR to determine the contact parameters using FEM. The ball-on-socket model developed for FEM which represents the acetabular cup and femoral head tribo-pair is shown in Fig.2. The cup and head are placed with a 45° inclination angle which is the standard inclination angle between the pelvis-femur contact [6]. The material properties and part geometries given to the FEM model are already described in Table 1 and 2 respectively. ABAQUS software is used to carry out the FEM simulations. The pre-processing conditions for the FEM simulations regarding the mesh type, contact interactions and application of boundary conditions are similar to the published literature [1].

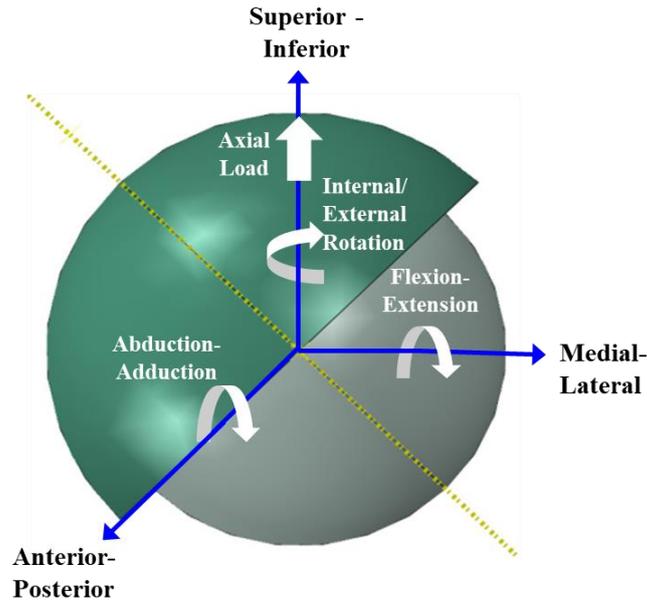

**Fig. 2.** FEM model for hip implants representing the cup and head tribo-pair

Since, both the analytical models meet the frictionless criteria, frictionless contact is taken to be the tangential feature for the FEM analysis. Furthermore, as established in the literature [42], the coefficient of friction applied in the contact region has very little impact on the estimate of the contact pressure which is used for the wear prediction. Considering friction will also increase the computational time in FEM as reported.

Mesh convergence study is performed using FEM based on the maximum contact pressure condition criteria. Through this study, it is found that 0.5 mm mesh is the converged mesh size irrespective of different clearances, head size and thicknesses. The comprehensive results i.e., maximum contact pressure, contact pressure distribution, contact radius and maximum deformation from the FEM for the converged mesh size are considered to verify with the existing analytical models.

**2.5 Dataset from FEM and ANN model**

Researchers discovered the concept of neural networks from the human neurons and their functionality by training them from childhood. Similar way, neural networks are trained from the data set and predict the values accurately. These neural networks have interconnected neurons organised in layers. In these, hyperparameters configuration plays a vital role in the training process and prediction. Hyperparameters such as the number of layers and neurons, learning rate, batch size, activation functions, and regularization techniques given by the user significantly affect training process.

The hyperparameters are determined using frequently used optimization techniques such as Grid Search, Random Search, and Bayesian optimization. The raw dataset mostly includes a lot of non-linearity and complexity, to deal this, activation functions come handy by inducing non-linearity into the network. The hyperbolic tangent function (tanh) is a commonly used activation function. This function maps the given input data to a range between -1 and +1, which exhibits like an S-shaped curve centred on 0, which means the outputs have zero mean, otherwise no effect at all. This property is useful with training data and improve the convergence.

Optimisers used to modify the model's parameters during training. They reduce the difference between predicted and target outputs by updating parameters based on gradients of the loss function. To maintain adaptive learning rates for every network parameter, the Adam (adaptive moment estimation) optimizer blends momentum-based and adaptive learning rate techniques. It effectively updates the model parameters during training by combining the adaptive learning rates and momentum, which promotes not only quicker convergence but also improved performance.

In the present study, the optimal hyperparameters are obtained through the Grid search method. Grid search is a hyperparameter optimization method that exhaustively searches through a specified set of hyperparameter values for an estimator. It is a brute-force method, meaning that it tries all possible combinations of hyperparameter values which could be computationally expensive if the number of hyperparameters were large. The search space parameters which are used for optimising the neural network are shown in Table 3.

**Table 3**

Search space parameters used for optimising the neural network

| Network specification | Search space |
| --- | --- |
| Number of hidden layers | 2 to 4 |
| Number of neurons per layer | 16 to 128 |
| Activation function | Sigmoid/ Hyperbolic tangent/ Rectified linear unit |
| Optimiser | Adam/ Stochastic gradient descent |
| Batch size | 16 to 64 |
| Learning rate | 0.1, 0.01, 0.001 |
| Epochs | 200 to 1000 |

The dataset from FEM has output contact values such as maximum contact pressure, maximum contact radius and maximum elastic deformation from the FEM for the respective input conditions discussed in section 2.3. This dataset is used to train the ANN model. The accuracy of the ANN model is quantified with the help of coefficient of correlation ($R^2$) and mean square error (MSE) shown in Eqs. (15) and (16) respectively.

$$R^2 = 1 - \frac{\sum_{i=1}^{N}(G^{actual} - G^{predict})^2}{\sum_{i=1}^{N}(G^{actual} - G^{mean})^2} \qquad (15)$$

$$MSE = \frac{1}{N}\sum_{i=1}^{N}(G^{actual} - G^{predict})^2 \qquad (16)$$

where *G* refers to the considered output contact parameter and *N* refers to the number of samples in the test data.

As ML models involve complex decision-making processes, their ability for interpretation is sometimes questioned when they are applied. This problem is addressed by Shapley additive explanations (SHAP) analysis, which provides information on feature contributions and model outputs [43]. Shapley values obtained from the SHAP analysis gives appropriate weightage to each feature considering all potential combinations. The Shapley values also provide an understanding of their impact on model predictions regardless of the type of regression models. This analysis provides information which can be used to enhance the model predictability. The information from the analysis will be vital in determining the parameters which contribute the major effects on the output data. Hence, in this study SHAP analysis is used to determine the effect of individual parameters affecting the output contact conditions in the ANN model.

## 3. Results and Discussion

### 3.1 Maximum contact pressure - comparison between analytical models and FEM

This section will discuss the comparison of maximum contact pressure between the analytical models and FEM considering different input parameters as mentioned in section 2.3. As discussed earlier, the load cycle is split into 40 steps and the results are analysed against these 40 points between the FEM and analytical formulations. Figs. 3-5 show the comparison of maximum contact pressure over a single cycle between the FEM and Hertz models for different radial clearances and thicknesses considering 28 mm, 32 mm and 36 mm head sizes respectively in MoM tribo-pair. From Figs. 3-5, it is observed that the profile of the maximum contact pressure over a single cycle is similar to the input load cycle. The FEM contact pressure decreases when head size increases but increases when the radial clearance increases. Additionally, contact pressure slightly decreases as the thickness increases, independent of the head size and clearance.

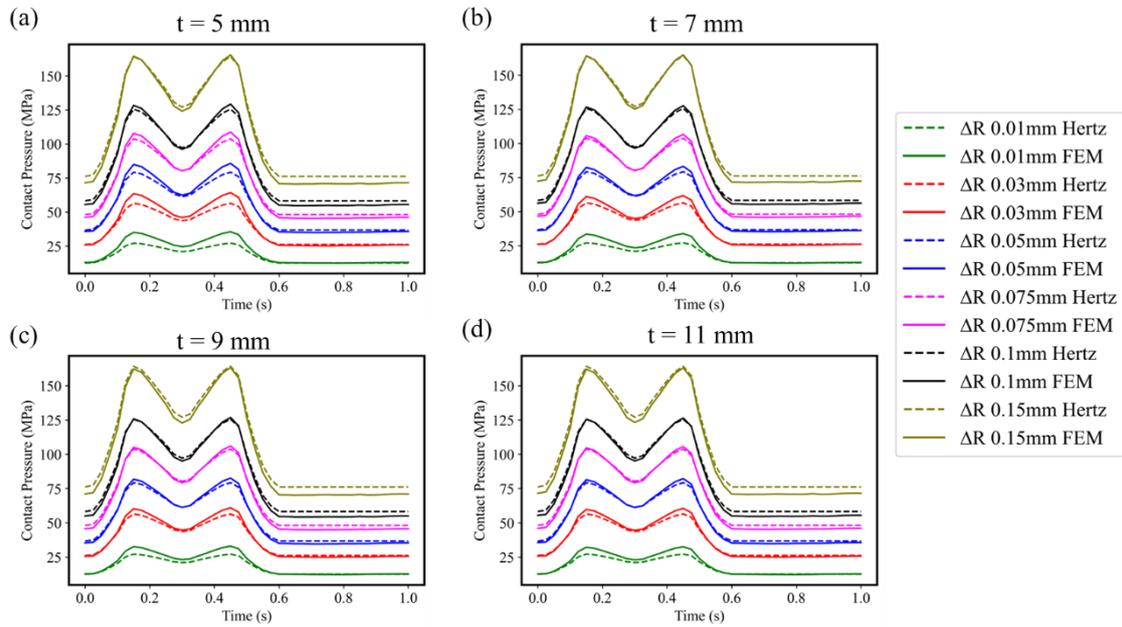

**Fig. 3.** Comparison of maximum contact pressure over a single cycle between FEM and Hertz analytical model for different radial clearances and thicknesses considering 28 mm head size in MoM tribo-pair

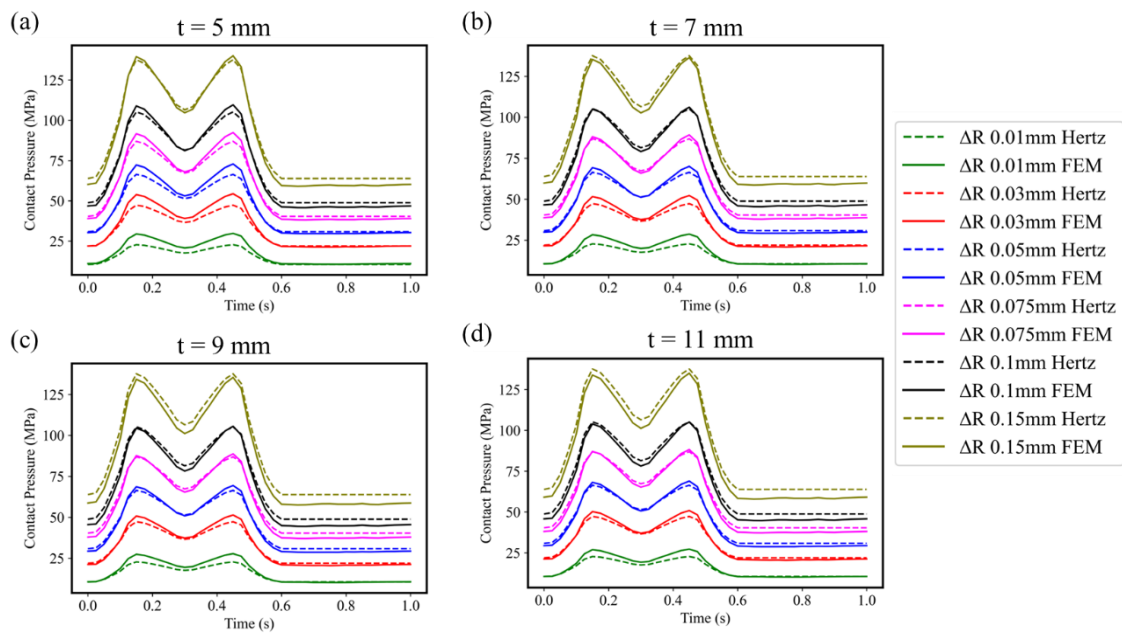

**Fig. 4.** Comparison of maximum contact pressure over a single cycle between FEM and Hertz analytical model for different radial clearances and thicknesses considering 32 mm head size in MoM tribo-pair

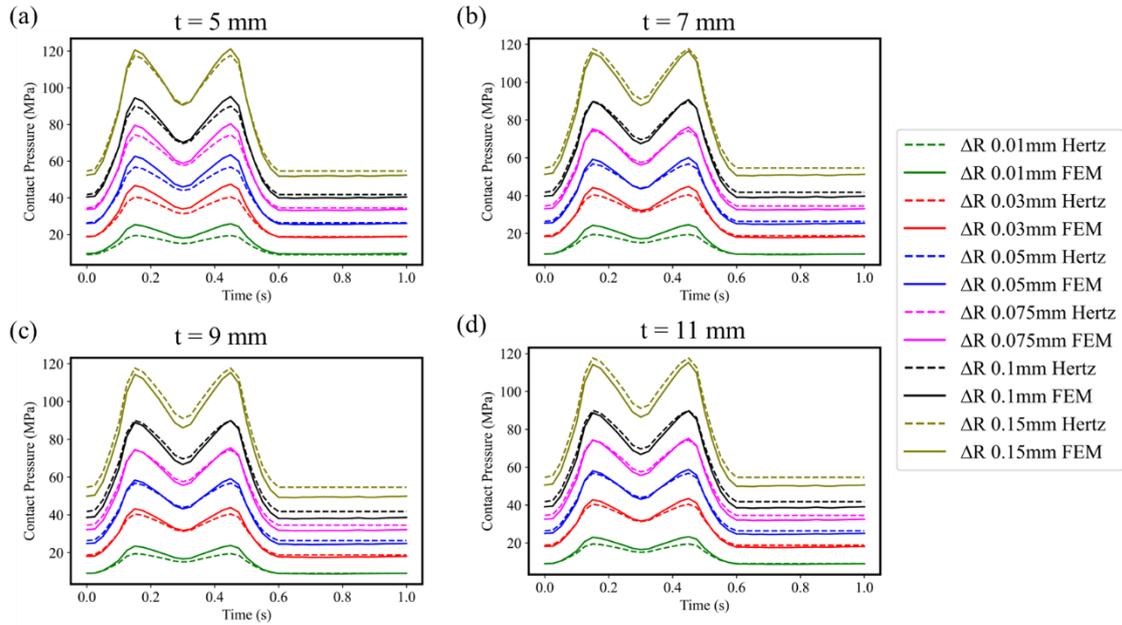

**Fig. 5**. Comparison of maximum contact pressure over a single cycle between FEM and Hertz analytical model for different radial clearances and thicknesses considering 36 mm head size in MoM tribo-pair

      Qualitatively, it is noted that in certain instances, the Hertz model agrees well with the FEM results, whereas in other instances, it does not. The maximum contact pressure results from the Hertz model vary significantly with the FEM results during the stance phase for the lower clearances, while in the swing phase for the higher clearances independent of the head size and thickness. Even though the contact pressure is less during the swing phase and the results of the considered analytical model doesn't agree well with FEM, this phase should not be ignored as it occupies 40% of the single cycle. The maximum contact pressure results from the FEM are largely underestimated during the stance phase but overestimated during the swing phase. This signifies clearly that the Hertz model has a specific zone of prediction capability.

      Similar to the above comparison approach, the maximum contact pressure results over a single cycle between the FEM and Fang models for different radial clearances and thicknesses considering 28 mm, 32 mm and 36 mm head sizes respectively in MoM tribo-pair are shown in Figs. 6-8.

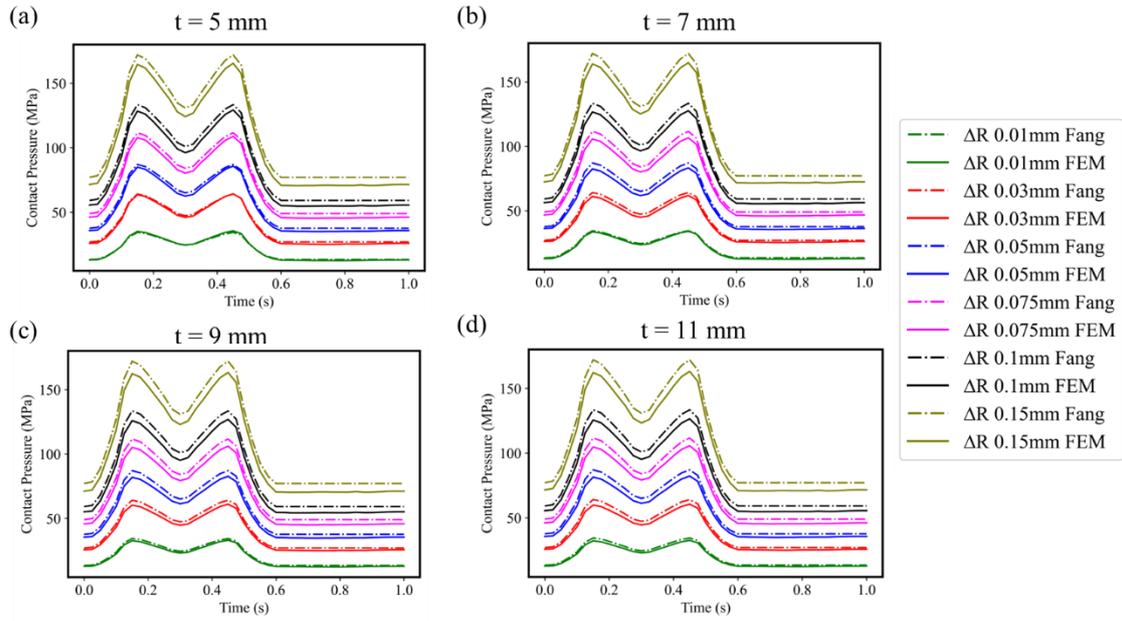

**Fig. 6.** Comparison of maximum contact pressure over a single cycle between FEM and Fang analytical model for different radial clearances and thicknesses considering 28 mm head size in MoM tribo-pair

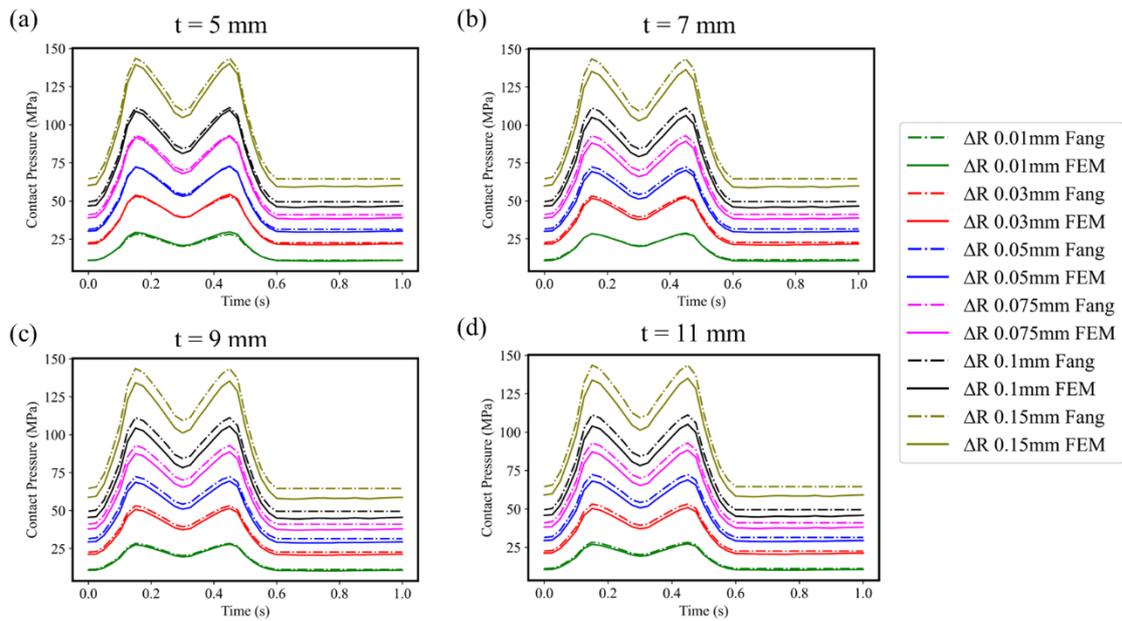

**Fig. 7.** Comparison of maximum contact pressure over a single cycle between FEM and Fang analytical model for different radial clearances and thicknesses considering 32 mm head size in MoM tribo-pair

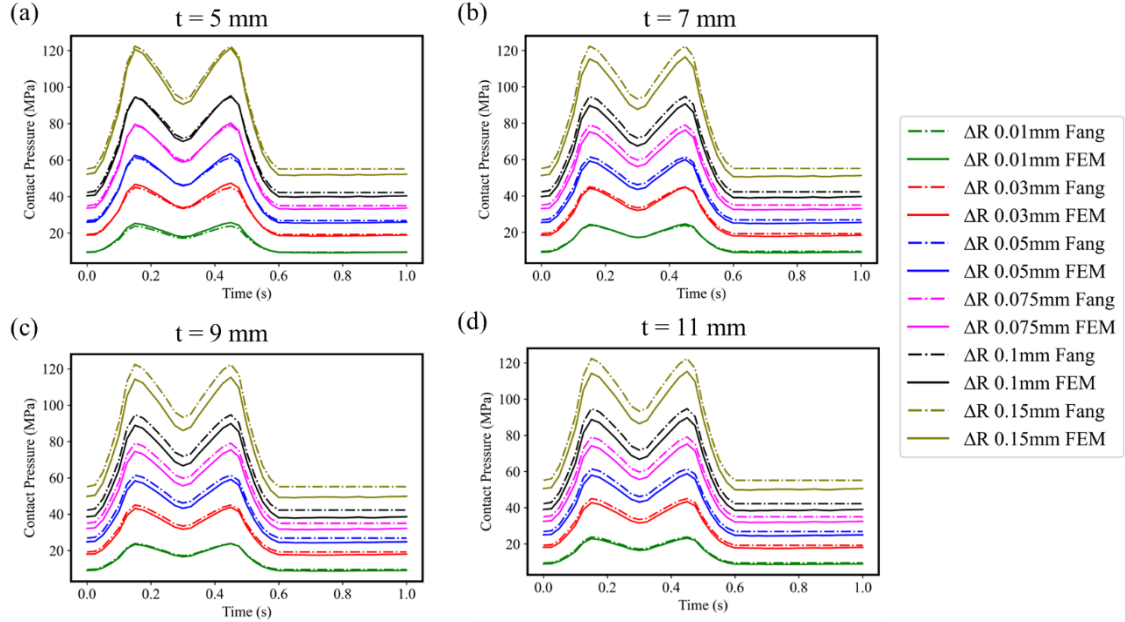

**Fig. 8.** Comparison of maximum contact pressure over a single cycle between FEM and Fang analytical model for different radial clearances and thicknesses considering 36 mm head size in MoM tribo-pair

In contrast to the Hertz formulation, it is observed qualitatively from Figs. 6-8 that the Fang formulation agrees well with the FEM results at low clearance values, i.e., <50 μm throughout the gait cycle. But at high clearance values, the Fang model largely fails at both stance and the swing phases of the complete gait cycle. Another observation is, for all cases, Fang formulation slightly overestimates the maximum contact pressure with respect to FEM which is quite distinct from the Hertz formulation. The similar phenomenon is seen for the CoC tribo-pair when the maximum contact pressure results from the analytical models and FEM are compared; these results are given in Appendix A. For the comprehensive contact analysis, the analytical models should be able to predict the contact radius as well as the contact profile which will be discussed in the subsequent section.

**3.2 Contact radius and contact pressure distribution throughout the profile**

As discussed earlier in section 2, the contact between the cup and head, when estimated through FEM, appears like a circular profile during contact with a clearance as shown in Fig. 9(a). It is obvious from the Fig. 9(a) that maximum contact pressure occurs at the centre of contact. It is also evident from Fig. 9(a) that the contact pressure is not smooth at the edges due to numerical instability of FEM nodes. As the profile is axisymmetric, the contact radius is estimated as the maximum unit of coordinate axis measurement through the nodes from the maximum contact pressure (centre node) to the node where pressure reaches to zero as mentioned in Fig. 9(b).

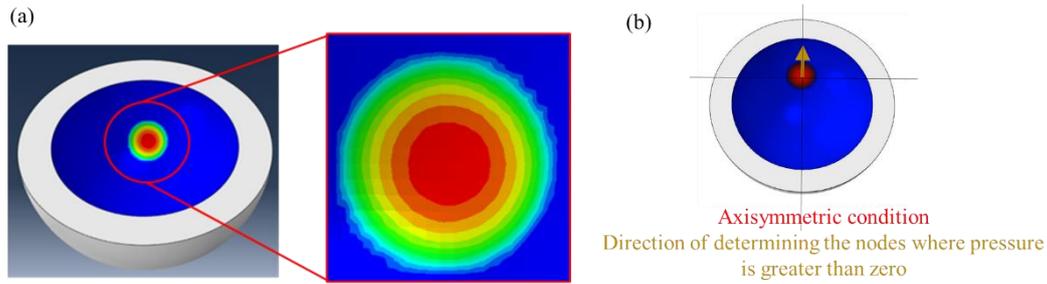

**Fig. 9.** (a) Typical circular contact profile distribution between cup and head on a given load by FEM. The edges of the circular profile in the magnified view denote the numerical instability (b) As the profile is axisymmetric, a set of nodes are taken from the FEM model for determining the profile along the arrow direction

To confirm the numerical instability, two different mesh sizes i.e. converged 0.5 mm and fine 0.1 mm are considered to explore further. The FEM contact profile is plotted axisymmetric for two different 10 μm and 150 μm radial clearances at extreme load cases, that is, 3000 N and 300 N for a particular 5 mm thickness considering MoM as shown in Fig. 10.

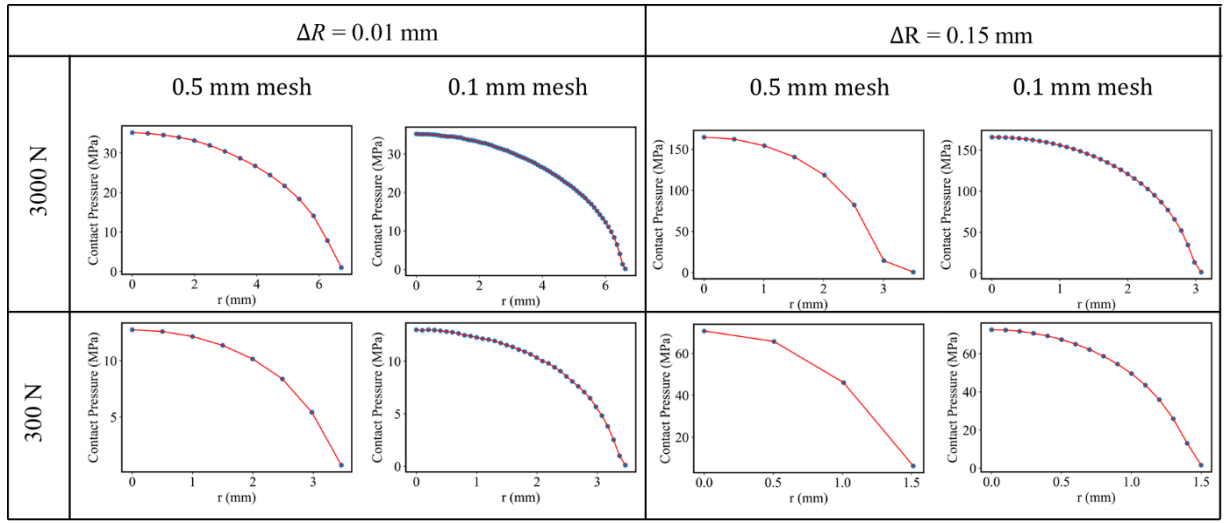

**Fig. 10.** FEM contact profile distribution for two extreme loads (3000 and 300 N) and radial clearances (0.01 and 0.15 mm) taken for MoM to study numerical instability considering 0.5 mm and 0.1 mm mesh. The blue dots denote the pressure data points which are connected using a red solid line to determine the profile.

Though the maximum contact pressure remains converged, it is observed from the Fig.10 that there is a significant variation in the contact radius which is evident explicitly at high clearance cases compared to low clearance cases for both the mesh sizes. This is due to the numerical instability at the edges of contact caused due to the mesh in FEM as shown in Fig. 9(a). This numerical stability due to the mesh can't be avoided completely but it decreases further as the mesh size reduces which captures the contact profile effectively. The 0.1 mm mesh gives a lot of data points, but at the cost of huge computational time compared to the 0.5 mm. Therefore, a 0.25 mm is chosen in between the selected mesh sizes to study only the contact radius and contact profile distribution as it gives sufficient data points even at high clearance cases. To predict the contact radius effectively from FEM 0.25 mm mesh, the best polynomial fit is ensured over the FEM data ignoring the last data point to estimate the correct contact radius and corresponding profile distribution as well. Based on the best n-polynomial fit to the data points from FEM, apart from 3000 N and 300 N, additionally one more load 1390 is taken as mid-load point represented as A, C and B respectively (as shown in Fig. 1). These points are taken for comparison of contact profile with FEM and analytical models which are shown in Figs. 11-13 considering MoM tribo-pair.

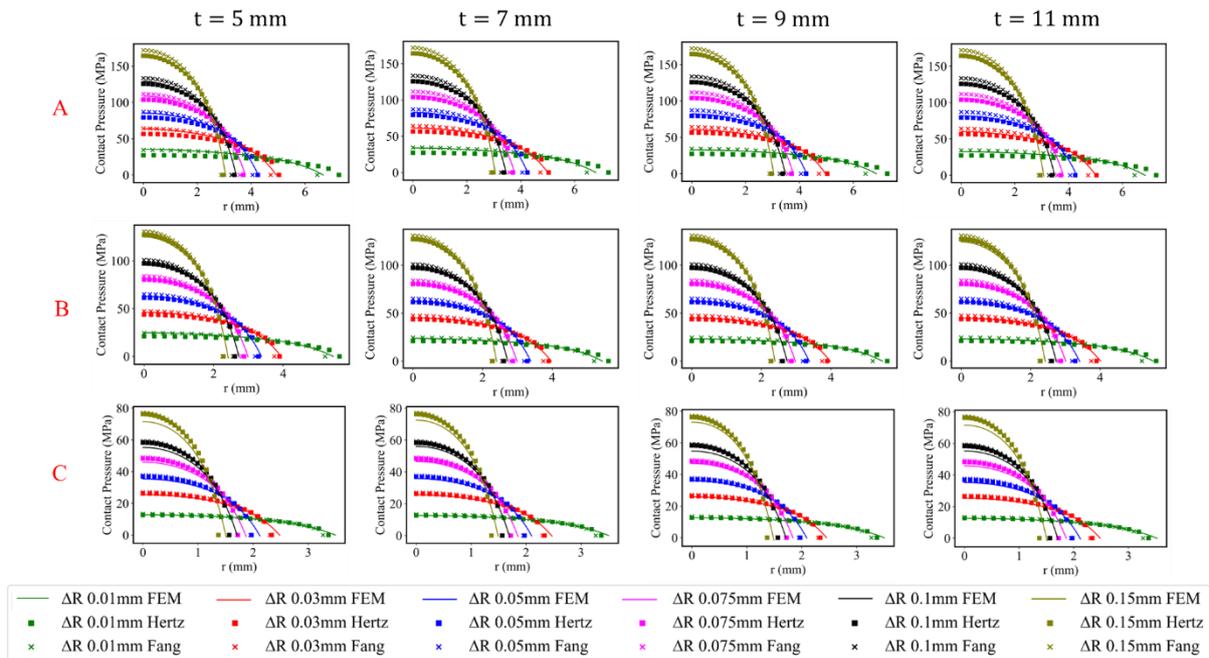

**Fig. 11.** Comparison of contact profile distribution between FEM, Hertz and Fang analytical formulation results at three different load points (shown in Fig. 1) for different radial clearances and thicknesses considering 28 mm head size in MoM tribo-pair

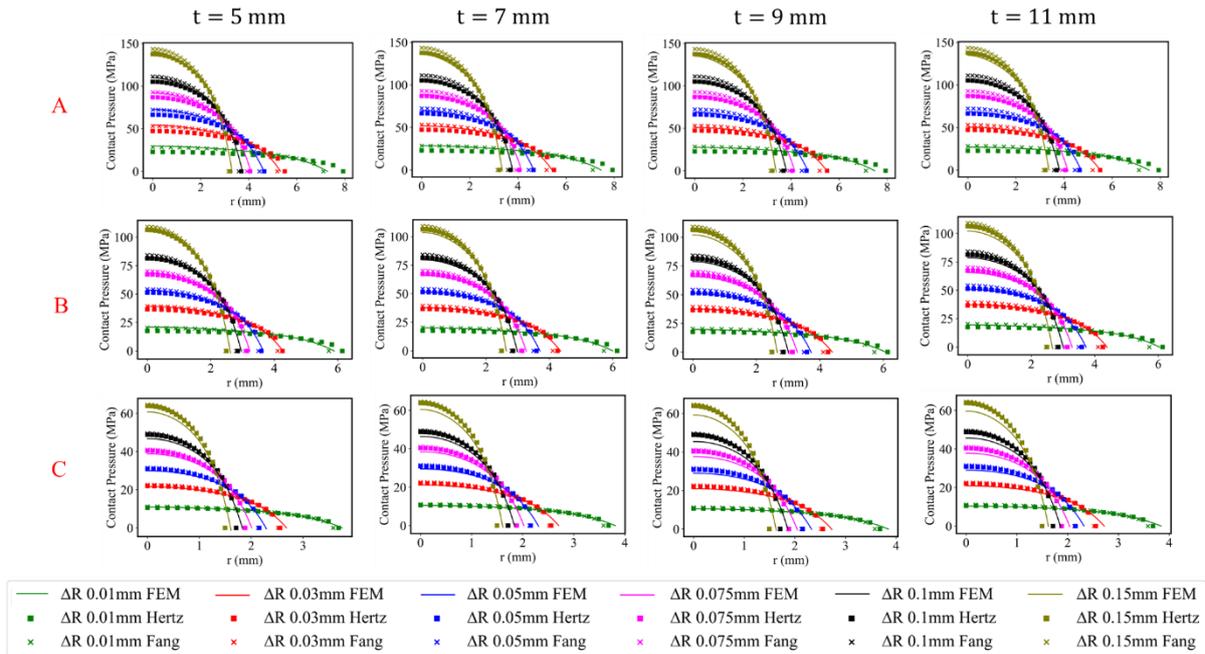

**Fig. 12.** Comparison of contact profile distribution between FEM, Hertz and Fang analytical formulation results at three different load points (shown in Fig. 1) for different radial clearances and thicknesses considering 32 mm head size in MoM tribo-pair

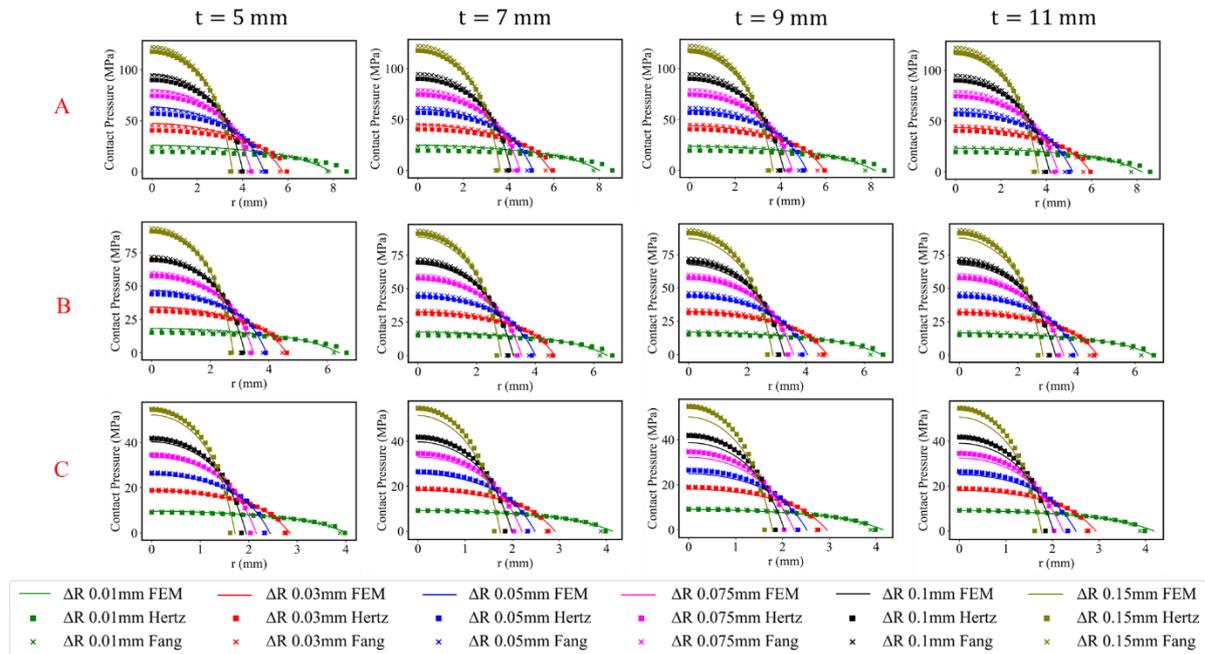

**Fig. 13.** Comparison of contact profile distribution between FEM, Hertz and Fang analytical formulation results at three different load points (shown in Fig. 1) for different radial clearances and thicknesses considering 36 mm head size in MoM tribo-pair

The difference in the maximum contact pressure between the FEM and analytical models is the same at the three load positions (A, B, and C), which are observed and reported in section 3.1. It is also inferred that there is a significant difference in the value of maximum contact radius in some load cases even though there is insignificant change in the maximum contact pressure between FEM and analytical models. Furthermore, a pressure change variation is still seen in some cases despite the maximum contact radius doesn't vary. The similar phenomenon is seen for the CoC tribo-pair when the contact radius results from the analytical model and FEM are compared; these results are given in Appendix B.

From the equations in section 2.1 and 2.2, it is evident that the maximum contact pressure is obtained in the analytical models using the contact radius *a* and pressure exponent factor *n* as well. Thus, for an accurate estimation of contact analysis, both *a* and *n* magnitudes must be precise. Interestingly, the magnitude of the exponent factor is obtained differently for both analyses. It is determined that in the Hertz model, the contact profile variation is assumed as parabolic, so *n* is fixed as 0.5. Whereas, in the Fang model, the pressure distribution exponent *n* is obtained from the curve fitting over the results of FEM simulations in steel-steel contacts. The value of *n* ranges from 0.5 to 0.26 in the Fang formulation but it depends on the CRP number. By means of curve fitting, Fang found that when the non-conformal contact approaches, *n* attains a maximum value of 0.5, which is equivalent to Hertz. Nevertheless, in the present study, for the same contact radius, the maximum contact pressure seems different in some load cases for both models. Clearly, this indicates that the *n* value maybe different for hip implant conditions. Therefore, it is emphasised further that in order to do a comprehensive contact analysis, the pressure exponent and contact radius values must be determined correctly. Furthermore, these two corrected parameters pave the way to determine the elastic deformation more accurately.

**3.3 Maximum elastic deformation - comparison between analytical models and FEM**

Figs. 14 and 15 show the comparison of maximum elastic deformation throughout a cycle considering different radial clearances and thicknesses between FEM and analytical models i.e., Hertz and Fang respectively.

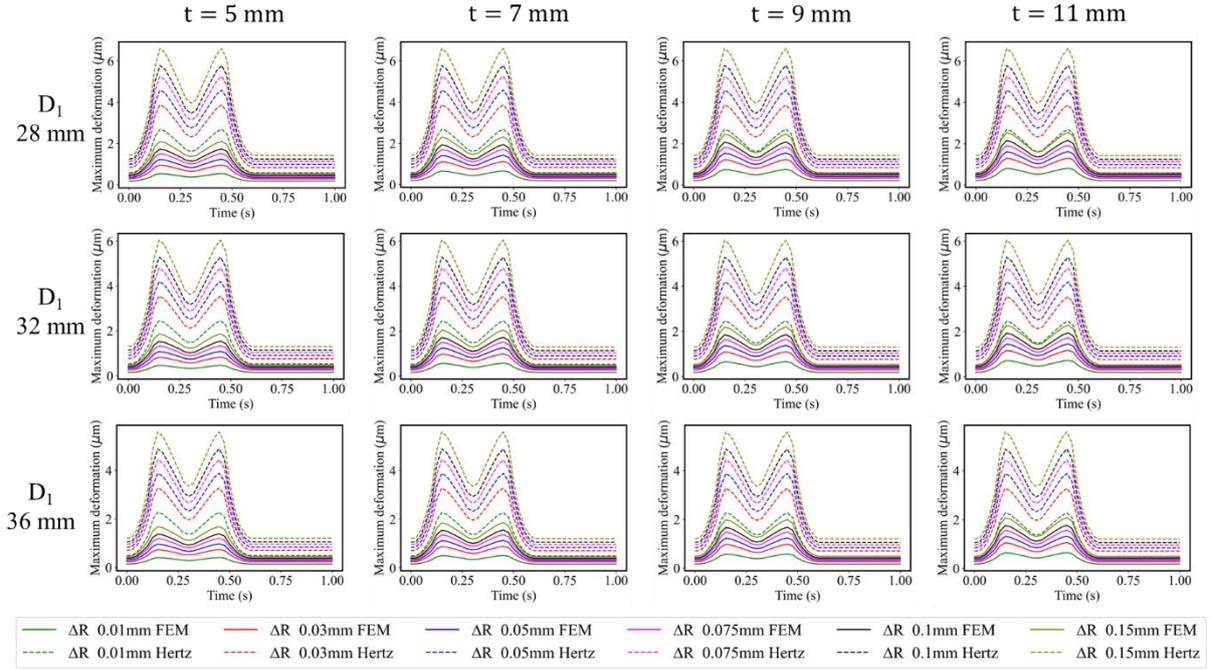

**Fig. 14.** Comparison of maximum elastic deformation over a single cycle between FEM and Hertz model for different radial clearances and thicknesses considering MoM tribo-pair

In many contact models, especially those involving elastic materials, the maximum penetration depth is typically associated with the maximum elastic deformation. Obviously, in elastic contact, the material deforms elastically up to a certain point before plastic deformation occurs. Therefore, the maximum penetration depth often coincides with the maximum elastic deformation. Similar to the maximum contact pressure, the profile of maximum elastic deformation is similar to the input load cycle as observed in Figs. 14 and 15. The maximum elastic deformation increases with acetabular cup thickness, according to the FEM results. As discussed earlier, the analytical models should also be able to estimate the maximum elastic deformation, which will be helpful in determining the possibility of an EHL regime in lubricated contacts. It is evident from Figs. 14 and 15 that both Hertz and Fang models completely differ from the FEM values for the maximum deformation considering all cases. As discussed earlier, the difference in maximum elastic deformation possibly due to the incorrect value of the estimated *a* and *n*. The value of maximum deformation depends on the coefficient *B*, which again depends on the value of *n* as a gamma function represented in Eq. (9). Further, the maximum elastic deformation obtained from the Hertz model is much larger than that obtained by the Fang model. This is due to the half-space non-conformal contact assumption, where the contact pressure is more focussed as a point rather than a surface assumption. In contrast, the Fang model integrates directly over the projected plane of the contact surface and then modifies the result to account for its effect of curvature on stress and deformation. Consequently, the magnitude of maximum elastic deformation in Fang is less than Hertz, but it is still high compared to the FEM results. The similar phenomenon is seen for the CoC tribo-pair when the maximum deformation results from the analytical model and FEM are compared; these results are given in Appendix C.

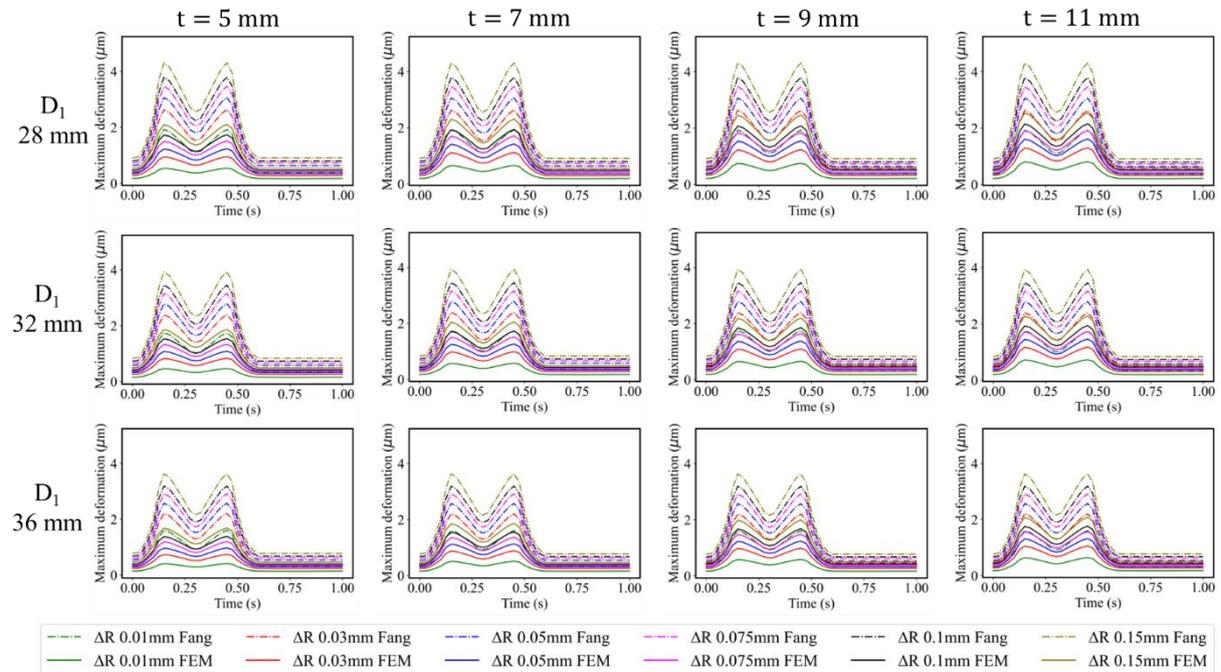

**Fig. 15.** Comparison of maximum elastic deformation over a single cycle between FEM and Fang model for different radial clearances and thicknesses considering MoM tribo-pair

Overall, through the qualitative analysis, it is inferred that both the analytical models disagree with the FEM contact results for all the possible input conditions considering a single gait cycle determined for THR. This implies that both models have an appropriate range of input conditions for which the models may be employed most effectively but not accurately. As known, the wear prediction can be carried out with the same accuracy as numerical models while saving the computational time by using these analytical models. However, wear prediction involves from a million to several million cycles depending on the application, and cascading effect plays a major role in it. Therefore, to avoid the cascading error accumulation in the wear prediction, the analytical models should have high accuracy.

Further, for comprehensive contact and wear analysis, the value of contact pressure and profile distribution should be accurate for better and accurate prediction of wear. Additionally, the determination of the correct value of maximum elastic deformation would help in finding EHL film thickness for the operating conditions of hip implants. At this juncture, both the analytical models fail to completely agree with the FEM conditions for the three output contact parameters considering all the input conditions. Furthermore, it's critical to perform a quantitative analysis and identify the specific reason for the unpredictability of existing analytical models in relation to the FEM results.

**3.4 Quantitative analysis of contact conditions for the analytical models**

To analyse the contact conditions quantitatively, specifically for hip implants, a new dimensionless parameter is introduced and named Hip Contact Radius Proportion (HCRP). It is defined as the ratio of the FEM contact radius to the femoral head radius, which is described in Eq. (18).

$$\text{HCRP} = \frac{a}{R_1} \qquad (18)$$

Fang [25] proposed a dimensionless parameter CRP, which was defined and fixed according to the socket radius. In hip implant tribo-pair, the head size is fixed, and acetabular cup size is varied according to the manufacturer taking the radial clearance into account. Therefore, the dimensionless parameter HCRP is defined according to the femoral head radius in the present study. The HCRP is used as a measurement to compare the values of output contact parameters between the analytical models and FEM. As standard industrial practice, the acceptance relative error between the FEM and analytical model is kept at 5%. From the load cycle, 23 distinct points are taken combined with 6 radial clearances, 4 thicknesses, 3 head sizes, and 2 material tribo-pairs contribute to 3312 input data points. The HCRP is obtained for all the 3312 data points and used as a measurement over the X-axis to check against the relative errors of analytical models against FEM for the chosen three output contact parameters. The HCRP value varies from 0.08 to 0.496, considering all the input parameters in the present study. The smaller the HCRP value denotes non-conformal contacts, while the larger the HCRP value represents conformal contacts. Fig. 16(a) and 16(b) show the relative error of maximum contact pressure between the FEM and analytical models, i.e., Hertz and Fang, respectively, over the calculated HCRP. Similarly, Fig. 16(c) and 16(d) show the relative error of maximum contact radius between the FEM and analytical models over the calculated HCRP.

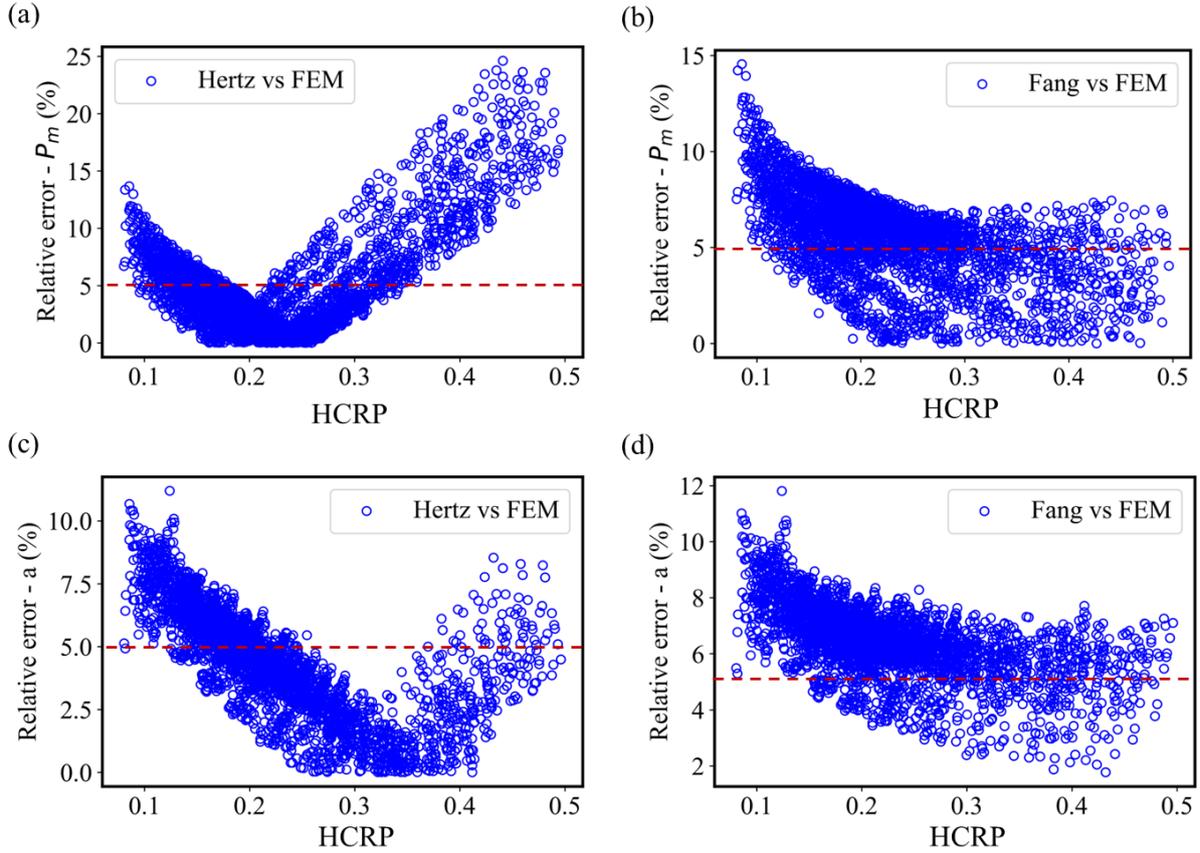

**Fig. 16.** Relative error distribution between analytical models and FEM (a) Hertz vs FEM for maximum contact pressure (b) Fang vs FEM for maximum contact pressure (c) Hertz vs FEM for contact radius (d) Fang vs FEM for contact radius

It is observed from Figs.16(a) and (b) that the maximum relative error of Hertz model (25%) is higher than the Fang model (15%) in determining the maximum contact pressure. Nonetheless, the overall relative error in the maximum contact radius is almost the same for both models as observed from Figs.16(c) and (d). Furthermore, considering the Hertz model alone at HCRP more than 0.3, even if the relative error in contact radius is less, the relative error in maximum contact pressure is high at this zone. Contrastingly in the Fang model, the relative error in maximum contact pressure is less even though the relative error in contact radius almost remains same as that of Hertz. It should be noted that the relative error denotes the absolute difference of magnitudes between the two output variables. To understand the importance of $n$, a particular case of pressure distribution for 3000 N load considering 9 mm thickness for a 28 mm MoM tribo-pair head is computed for the analytical models and plotted against FEM in Fig.17. From Fig.17, it is inferred that the contact radius predicted by Fang is lesser than FEM, at the same time it overestimates the contact pressure than FEM. While, Hertz predicts a higher contact radius than FEM which underestimates the contact pressure.

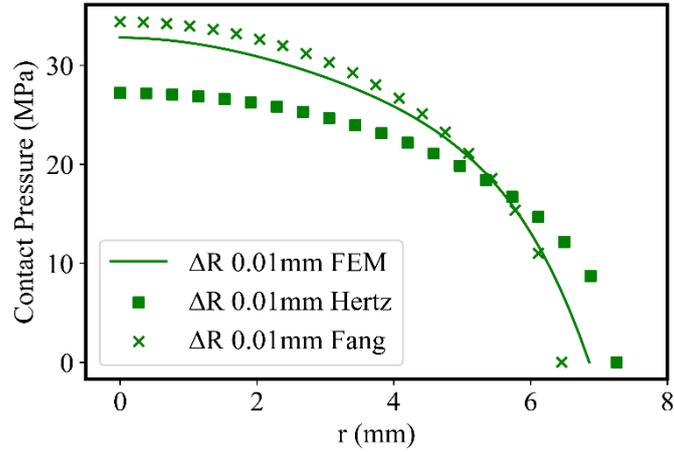

**Fig. 17.** Contact pressure distribution at 3000 N load for 9 mm thickness for 28 mm MoM tribo-pair head

It is important to understand that the absolute relative difference of these contact radius values is same, but the exponent *n* determines the maximum pressure as per the profile variation assumption. Fang determined that the value of *n* is constant at 0.5 when CRP varies from 0 to 0.6 [25]. The CRP is also estimated in the present study considering all the input parameters. The maximum value of CRP found out is less than 0.6 in this study, which makes the *n* to set always at 0.5.

It should be recalled that *n* finalises the maximum contact pressure and profile distribution based on the contact radius, as was previously discussed in section 3.2. To accommodate the pressure profile, the value of n as 0.5 causes a high relative error with contact pressure in overestimated contact radius compared to underestimated contact radius as observed considering a particular case as shown in Fig. 17. However, considering n as 0.5 might cause relative error in contact pressure even if the radius remains same for both analytical models and FEM as the profile variation is assumed to be parabolic. This hints that the consideration of *n* value as 0.5 is incorrect for hard-on-hard hip implants throughout the gait cycle considering all the input parameters.

It is vital to explicitly find out the effect of individual input parameters, which affect the relative error in the contact parameters. Considering only the Hertz model, Figs. 16(a) and 16(c) are updated with the five input individual parameters (radial clearance, thickness, femoral head size, applied gait load and material tribo-pair) represented in Fig.18. For better understanding and visualisation aspect, the load, *W,* is classified as low load range (300-1200 N), mid load range (1200-2100 N) and high load range (2100-3000 N).

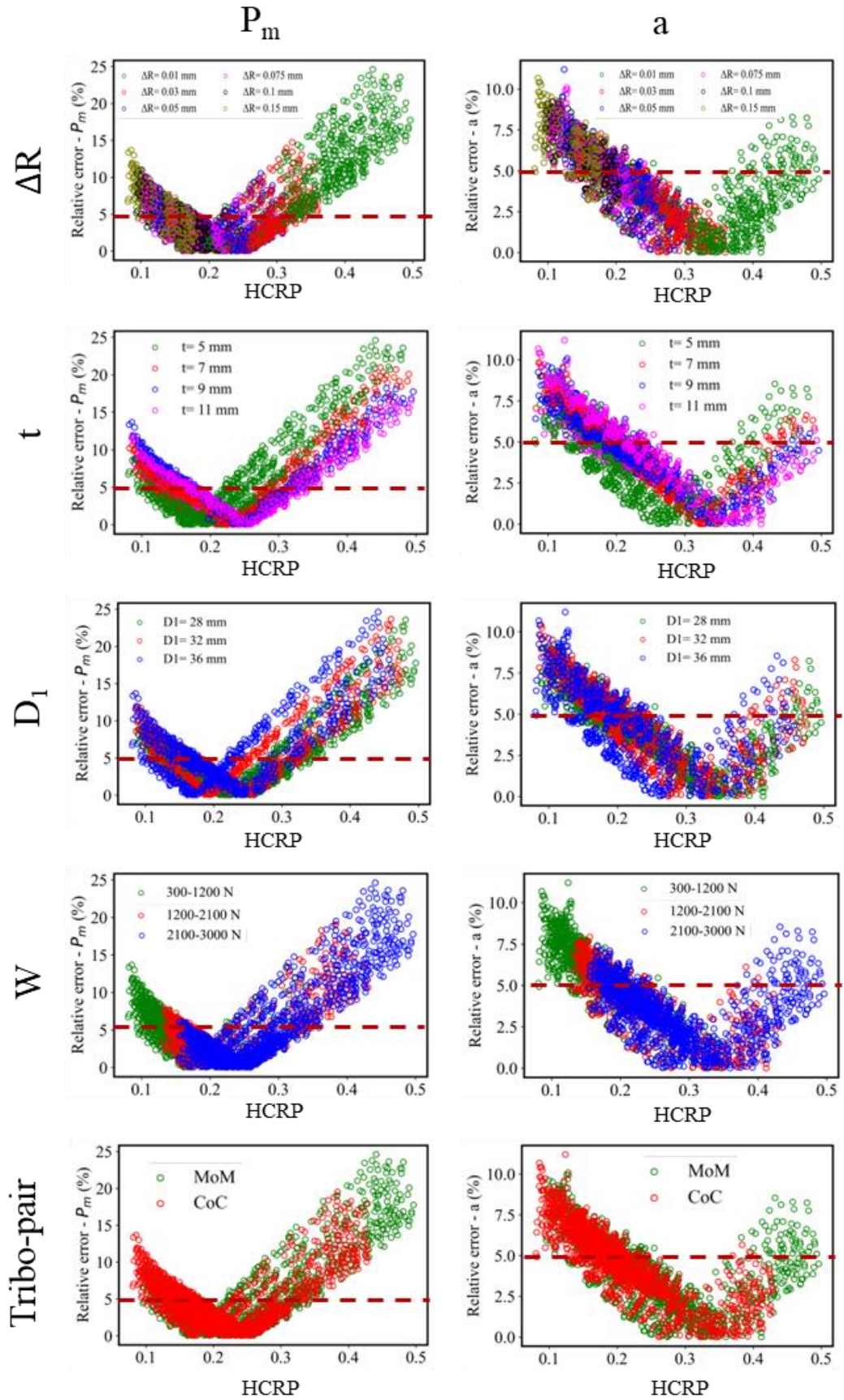

**Fig. 18.** Relative error of maximum contact pressure and contact radius between FEM and Hertz considering (a) clearance (b) thickness (c) femoral head size (d) applied load and (e) material tribo-pair.

From Fig.18, it is observed that for both maximum contact pressure and contact radius, it is evident again that the Hertz model fails mostly at both low load-high clearance and high load-low clearance conditions irrespective of the other input parameters. Comparing the contact pressure and contact radius considering $\Delta R$ and $W$, the relative error of contact radius is less for high load-low clearance compared to low load-high clearance. However, the relative error in maximum contact pressure is very high for high load-low clearance due to conformal contact. This is because the more conformal the contact becomes, the Hertz model fails as it violates the half-space approximation which is also discussed in [25,44]. Although, the effect of thickness is less, the relative error slightly increases or decreases in both pressure and radius due to changes in thickness. There is not much effect on the head size in the relative error. Further within the high load-low clearance condition, the error in MoM is higher compared to CoC, which indicates that the contact is more conformal in the former than the latter. This is due to the lower stiffness of metal compared to ceramics.

Similar to Hertz vs FEM comparison, Fig.19 illustrates how the individual parameters affect the relative error between Fang and FEM models derived from Figs. 16(b) and 16(d). Contrasting to the Hertz model, the relative error for both pressure and radius in Fang model increases as the clearance increases. The effect of thickness and head size variation on the relative error is similar to the Hertz model for both pressure and radius. Interestingly, the effect of material tribo-pair is insignificant as Fang accounted for both conformal and non-conformal approximations, i.e., deriving a new model in addition to the assumptions of Hertz.

The above discussions clearly indicate that both the models fail to predict the contact conditions considering all the input parameters for hard-on-hard hip implants. The contact problems are highly nonlinear due to the geometry, load and status of contact (nodes come in and out of contact) in hip implants [25]. The relationships can be directly or inversely proportional between the input and output parameters but the magnitude of proportionality throughout the gait cycle will vary non-linearly. For example, the contact radius increases rapidly with load, specifically when the load is in low load range. Further, when the load reaches past the mid load range, the contact radius slowly increases to a limiting value. Henceforth, determining the actual relationships explicitly over a large range of geometrical and operational conditions is arduous using analytical models.

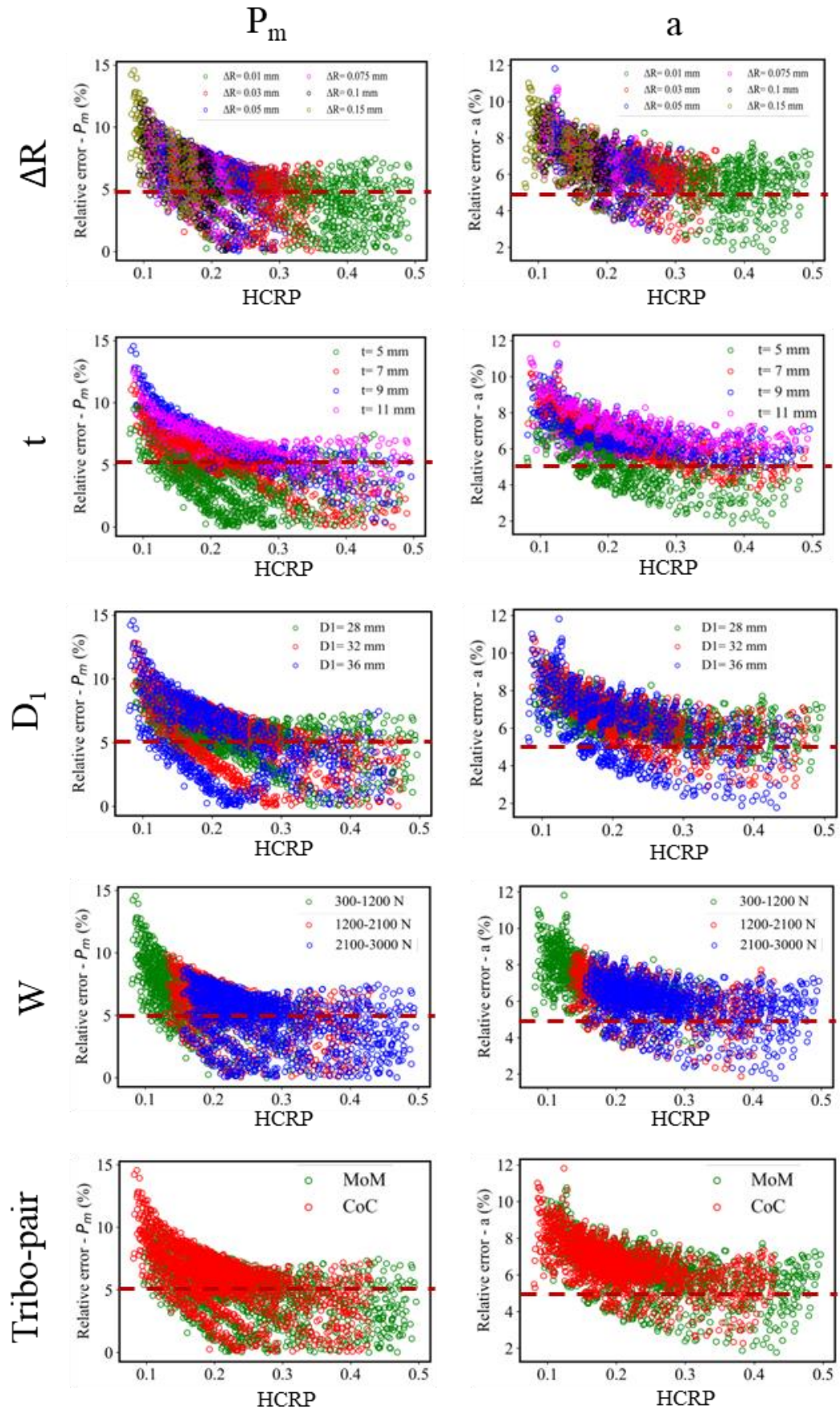

**Fig. 19.** Relative error of maximum contact pressure and contact radius between FEM and Fang considering (a) clearance (b) thickness (c) femoral head size (d) applied load and (e) material tribo-pair.

**3.5 Actual pressure distribution exponent *n* for hip implant conditions from FEM**

As discussed in the previous sections, the contact pressure distribution, contact radius and maximum elastic deformation depends on the value of *n* as shown in the Eq. (12). It is important to determine the correct value of *n* particularly for the hard-on-hard hip implants for accurate contact analysis and subsequently for wear analysis. As reported [25], the value of *n* causes the pressure distribution over the circular boundary profile to vary like a parabola if it is equal to 0.5. It is well known that the Hertz formulation always follows the pressure distribution as a parabola, when *n* is fixed to 0.5. The Fang model proposes [25] to keep *n* as 0.5 up to a CRP of 0.6, beyond which the *n* decreases up to 0.26 when CRP reaches to 1. To estimate the correct distribution of pressure for hip implant conditions, the pressure profile Eq. (12) is varied with different *n* values and fit with the FEM contact pressure distribution of the 0.25 mm mesh. The best *n* value is taken as the fit having the minimum MSE. The *n* values of best fit profiles are plotted against HCRP and are shown in Fig. 20.

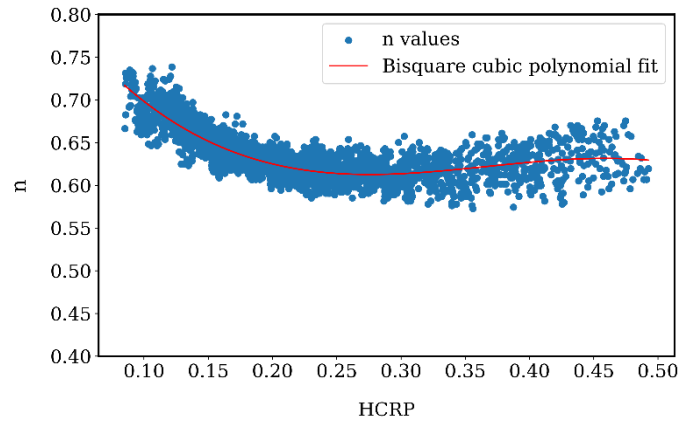

**Fig. 20.** Best pressure exponent n fit values with the FEM data. A bisquare cubic polynomial fit is performed over the data

The data points in Fig.20 are fitted using a bisquare weight cubic polynomial fit of non-linear regression analysis type having the MSE $2.13\times10^{-4}$. The pressure distribution exponent *n* equation using curve fit is described in Eq. (18).

$$n = 0.8699 - 2.3226\left(\frac{a}{R_1}\right) + 6.7688\left(\frac{a}{R_1}\right)^2 - 6.1406\left(\frac{a}{R_1}\right)^3 \quad (18)$$

By knowing the HCRP value, the correct value of *n* can be obtained using Eq. (18) which is valid only for hard-on-hard hip implants. From Fig.20, it is also observed that the *n* value in the range between 0.57 to 0.74 for the contact in hip implants. This clearly violates the findings of both Hertz and Fang models.

### 3.6 Data-driven ANN model

As discussed earlier, ANN models are used to implicitly determine the accurate relationships between the input and output parameters which are difficult to determine explicitly using the analytical models. Hence, the selected three output contact parameters, i.e., maximum contact pressure, contact radius value from n-polynomial fit and maximum deformation from FEM are taken as the dataset.

In order to investigate the effect of equivalent modulus, a third hip implant hard tribo-pair material, such as alumina (E = 413 GPa and $\upsilon$ = 0.235), is also taken into consideration. The FEM analysis is also carried out for the new material combination for all the input geometrical and operational cases. A dataset of 4968 data points obtained from FEM having 3 output parameters and 5 input parameters (section 2.3) is used to train the ANN model. As mentioned in section 2.5, the hyperparameters chosen for the ANN model are decided using the grid search optimisation. In machine learning, overfitting happens when a model fits the training data excessively, capturing valuable patterns and meaningless noise, leading to inaccurate predictions of new data. Some strategies to avoid overfitting in ANN are Dropout, L1 or L2 regularization and k-fold cross-validation technique. In the present study, a fivefold cross-validation technique is employed to avoid overfitting the data, as it splits the training dataset into five parts. The results are optimized based on the average score of the five sub datasets. The hyperparameters are optimized based on the best $R^2$ and low MSE scores considering all the output parameters and are tabulated in Table 4. The architecture of the optimised neural network model with the hyperparameters is shown in Fig. 21.

**Table 4**

Specification of optimised neural network

| Network specification | Optimal value |
| --- | --- |
| Number of hidden layers | 2 |
| Number of neurons per layer | 100 |
| Activation function | Hyperbolic tangent |
| Optimiser | Adam |
| Batch size | 64 |
| Learning rate | 0.001 |
| Epochs | 1000 |

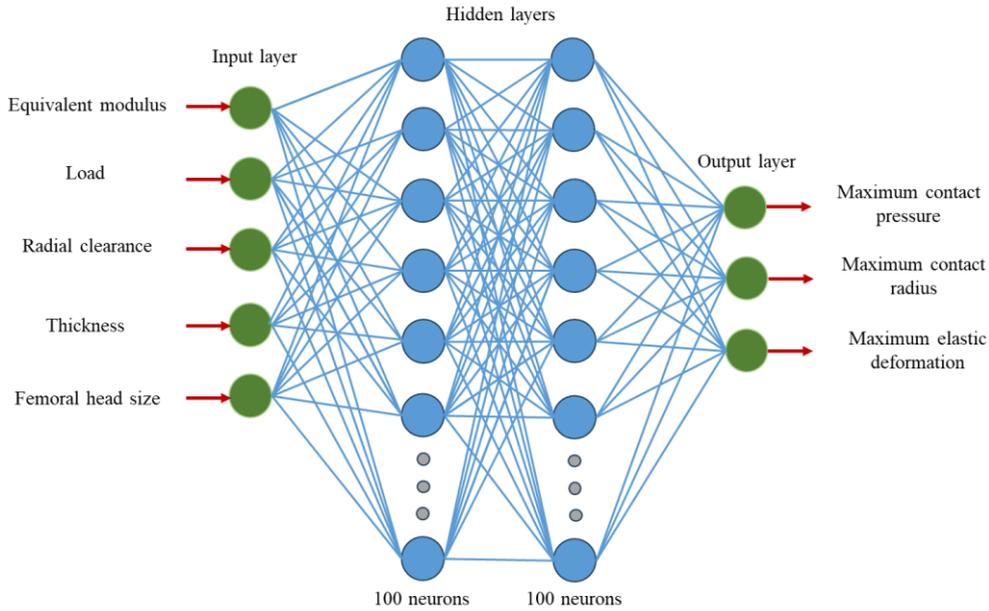

**Fig. 21.** Architecture of the optimised neural network

To check the accuracy of the ANN model considering all the three contact conditions, 20% of the input data set was used as the test data and the results are shown in Fig. 22. It is observed that the optimised ANN model using the grid search algorithm gives the minimum best coefficient of determination ($R^2 > 0.9914$) considering the three contact output parameters. This demonstrates the ANN model's reliability and stability in predicting the output contact parameters. Once trained, the computational time of the ANN model in predicting the three contact parameters for the input parameters in a single gait cycle is about 10 s. This is far less than the single FEM gait cycle, which takes about 67 minutes for the 0.5 mm mesh itself executed on a 64-bit Windows 10 Pro workstation with Intel Xeon processor at 3.3 GHz.

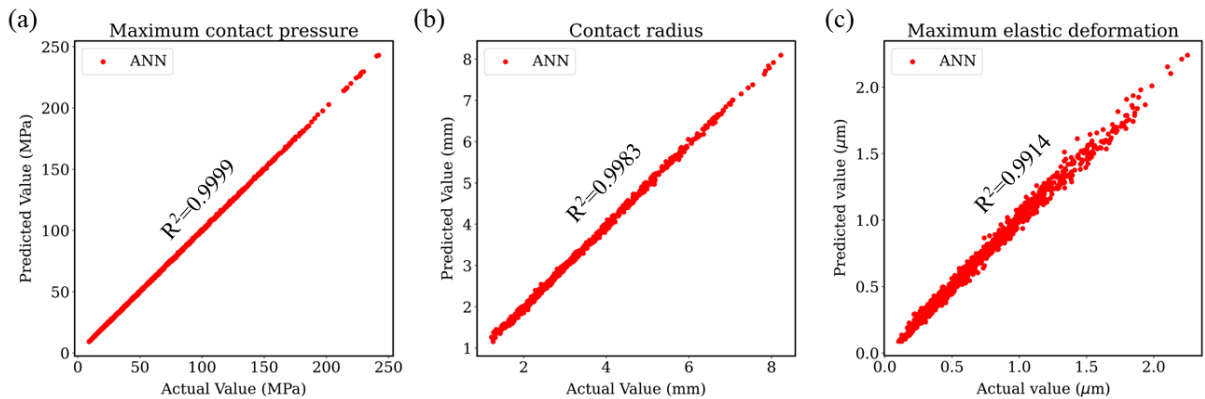

**Fig. 22.** $R^2$ performance of the ANN model in predicting the contact parameters using the data (a) contact pressure (b) contact radius and (c) maximum elastic deformation

## 3.7 Effect of individual input parameters affecting the contact parameters – SHAP analysis results

As discussed in section 2.5, the SHAP analysis is used to determine the percentage contribution of the individual input parameters, along with ranking as well. The results of the SHAP analysis for the respective output parameters are represented as a pie chart in Fig. 23. It is observed from Figs. 23(a) and 23(b) that radial clearance (≈45%) and thickness (≈1%) are the most and least influencing parameters respectively affecting the contact pressure and radius. The remaining input parameters arranged in-between are in the descending order: applied load, equivalent modulus, femoral head size. The acetabular cup thickness was not taken into account by either the Hertz or the Fang models, as discussed in sections 2.1 and 2.2. Though, the individual contribution of thickness is less for pressure and radius, it is enough to cause changes in relative error as inferred from qualitative analysis for both models.

Interestingly, as seen in Fig. 23(c), thickness has a significant contribution (≈8%) in the maximum elastic deformation, affecting the minimum film thickness and, consequently, the lubrication regime as well. Increase in the thickness of acetabular cup causes the contact force distributed over a wide area causing maximum deformation to increase. Unlike the contact pressure and radius, a small change in the thickness is expected to magnify the error in the deformation. Therefore, it is important and recommended to consider thickness in the analytical formulations, especially, for contact mechanics in hip implants.

This study is the first to report the percentage contribution along with the ranking of all features to the correlation output contact parameters. This enhances the understanding of ANN model behaviour and improves the analytical model interpretability which is not possible using numerical simulations. The contribution of the individual parameters in affecting output contact conditions will be helpful in developing a new analytical model particularly for hard-on-hard hip implants.

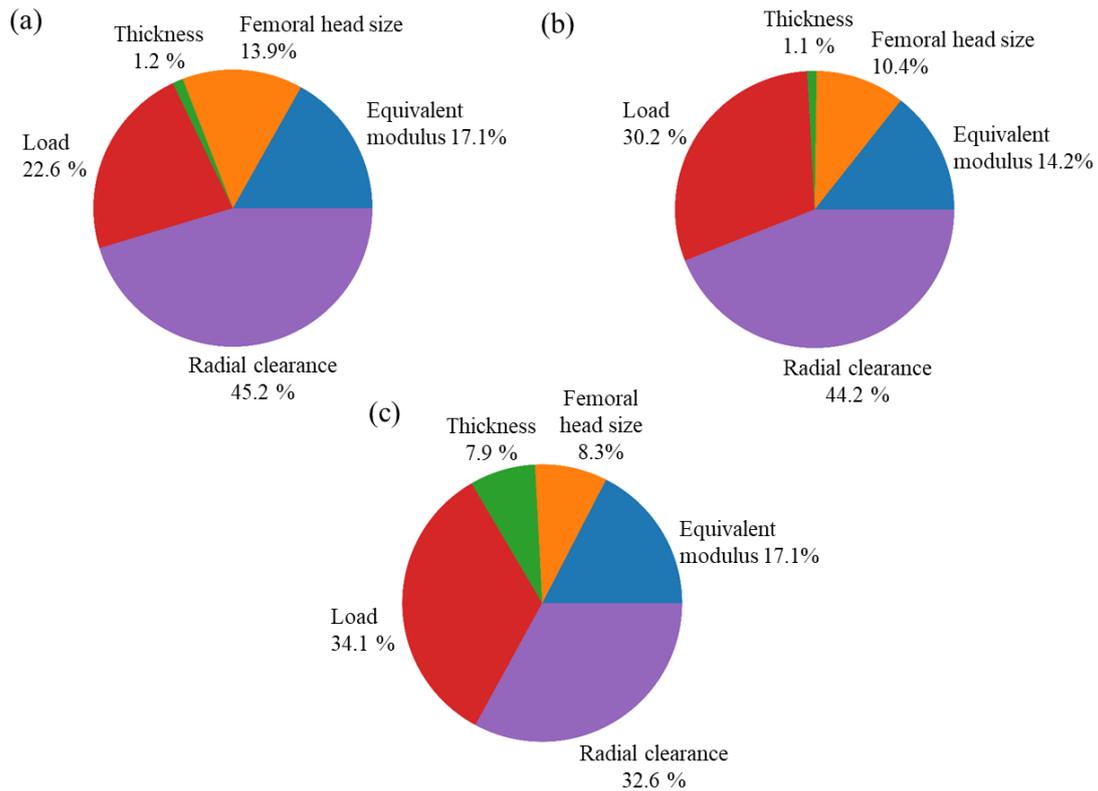

**Fig. 23.** SHAP analysis determining the effect of individual parameters affecting the output contact parameters (a) maximum contact pressure (b) contact radius and (c) maximum elastic deformation

In summary, the trained ANN model can predict the output contact parameters, i.e., maximum contact pressure, maximum contact radius and maximum elastic deformation with a high accuracy along with less computational cost than FEM. The data from the ANN model will help to determine the HCRP for a given input parameter condition. Further, the pressure distribution exponent $n$ is determined through Eq. (18), thereby giving the pressure distribution throughout the contact profile. The results from the contact output parameters using ANN may help in maintaining average contact pressure while starting in a simple flat-on-flat tribometer to simulate exact hip implant simulator conditions.

## 4. Conclusion

The present study examines the comprehensive contact analysis in hard-on-hard hip implants considering five input system parameters such as gait load, femoral head size, thickness of the acetabular cup, radial clearance and equivalent modulus of the tribo-pair. The contact conditions are analysed using existing analytical models (Hertz and Fang) as well as a data-driven neural network approach and validated with FEM. From this study, the following conclusions are given below:

- For a given complete gait cycle, both analytical models fail to predict the output contact parameters.
- For the quantitative analysis, a new dimensionless parameter HCRP is developed to quantify the prediction capability for the analytical models; the Fang model gives comparatively less error than the Hertz model.
- The SHAP analysis from the developed ANN model show that cup thickness has a significant contribution affecting the output contact conditions. Thus, it is recommended to consider cup thickness in the analytical model interpretation for hard-on-hard hip implants.
- A relationship is developed between pressure exponential $n$ and the HCRP. The maximum contact radius obtained from the developed ANN model can be used to determine the pressure distribution exponent $n$.
- Overall, a novel data-driven neural network model is developed which comprehensively predicts the contact conditions with higher accuracy and less computational cost specifically for hard-on-hard hip implants.

This study also has limitations of predicting only the initial conditions of the implant. However, as wear happens due to which clearance might change and the theoretical models must be enhanced to capture the change in geometry and giving corresponding contact pressures and profiles. Until now, FEM still remains as the prominent and widely used method for comprehensive wear analysis as the change in geometry of the parts is perfectly captured.

# Appendix A - Maximum contact pressure comparison between FEM and Analytical models for CoC

## Hertz formulation

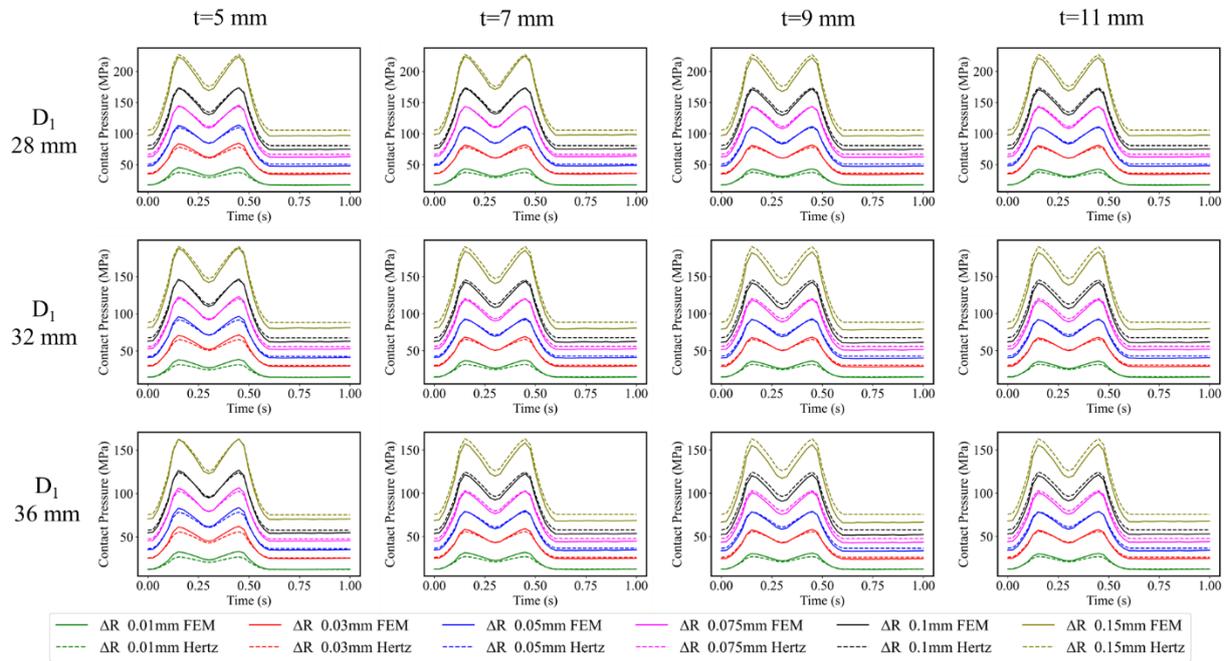

**Fig. A1.** Comparison of maximum contact pressure over a single cycle between FEM and Hertz analytical model for different radial clearances, head sizes and thicknesses considering CoC tribo-pair

## Fang formulation

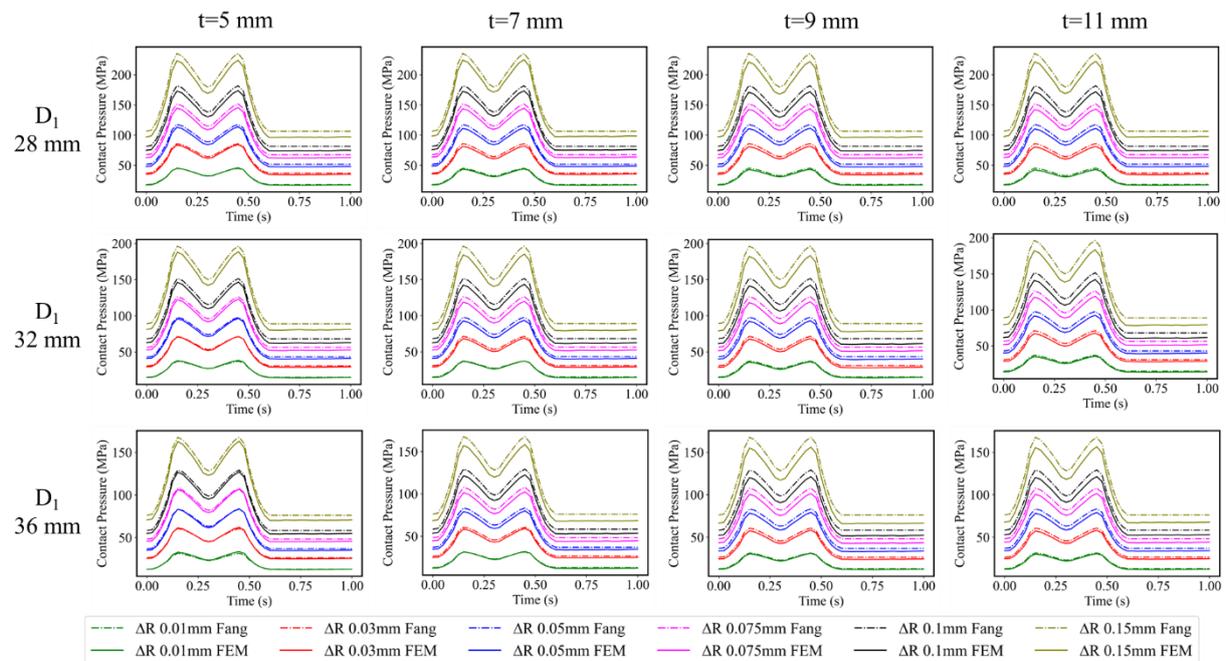

**Fig. A2.** Comparison of maximum contact pressure over a single cycle between FEM and Fang analytical model for different radial clearances, head sizes and thicknesses considering CoC tribo-pair

# Appendix B - Contact profile comparison between FEM and Analytical models for CoC

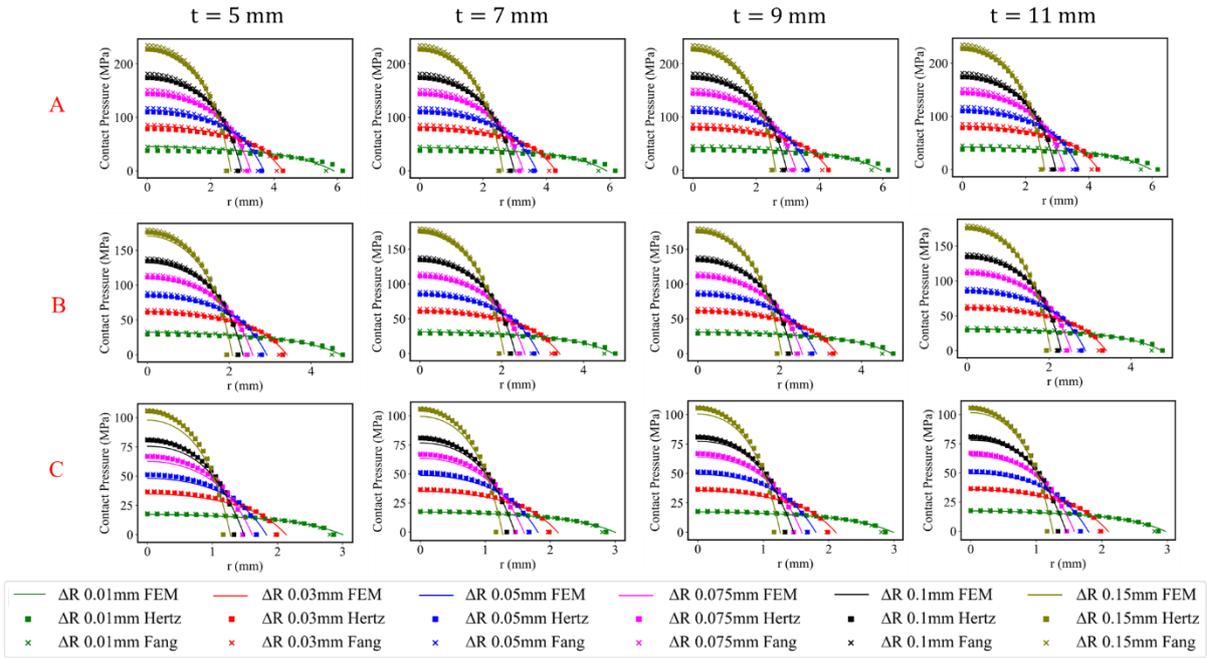

**Fig. B1.** Comparison of contact profile distribution between FEM, Hertz and Fang analytical formulation results at three different load points (shown in Fig. 1) for different radial clearances and thicknesses considering 28 mm head size in CoC tribo-pair

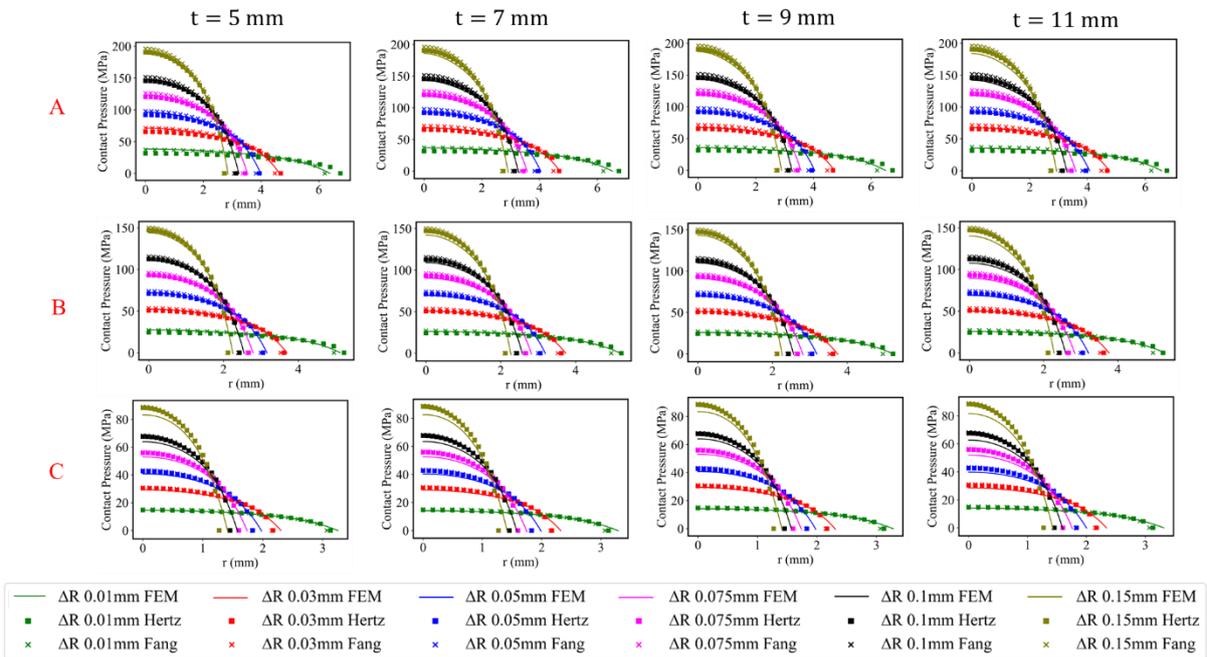

**Fig. B2.** Comparison of contact profile distribution between FEM, Hertz and Fang analytical formulation results at three different load points (shown in Fig. 1) for different radial clearances and thicknesses considering 32 mm head size in CoC tribo-pair

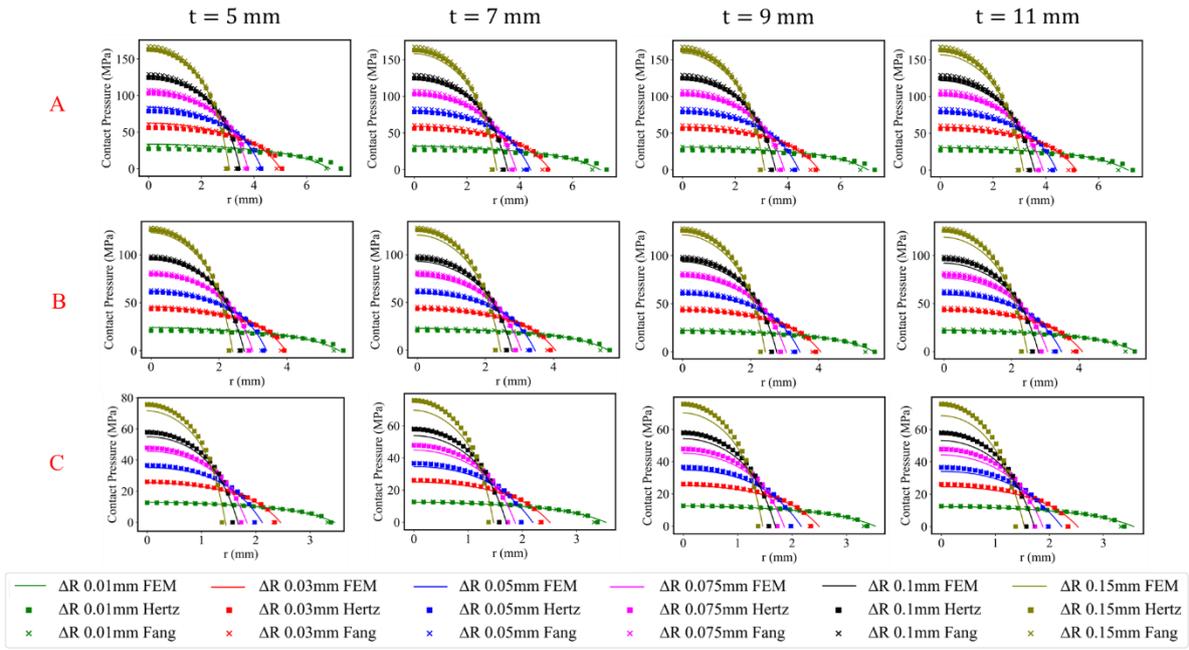

**Fig. B3.** Comparison of contact profile distribution between FEM, Hertz and Fang analytical formulation results at three different load points (shown in Fig. 1) for different radial clearances and thicknesses considering 36 mm head size in CoC tribo-pair

# Appendix C - Maximum elastic deformation comparison between FEM and Analytical models for CoC

## Hertz formulation

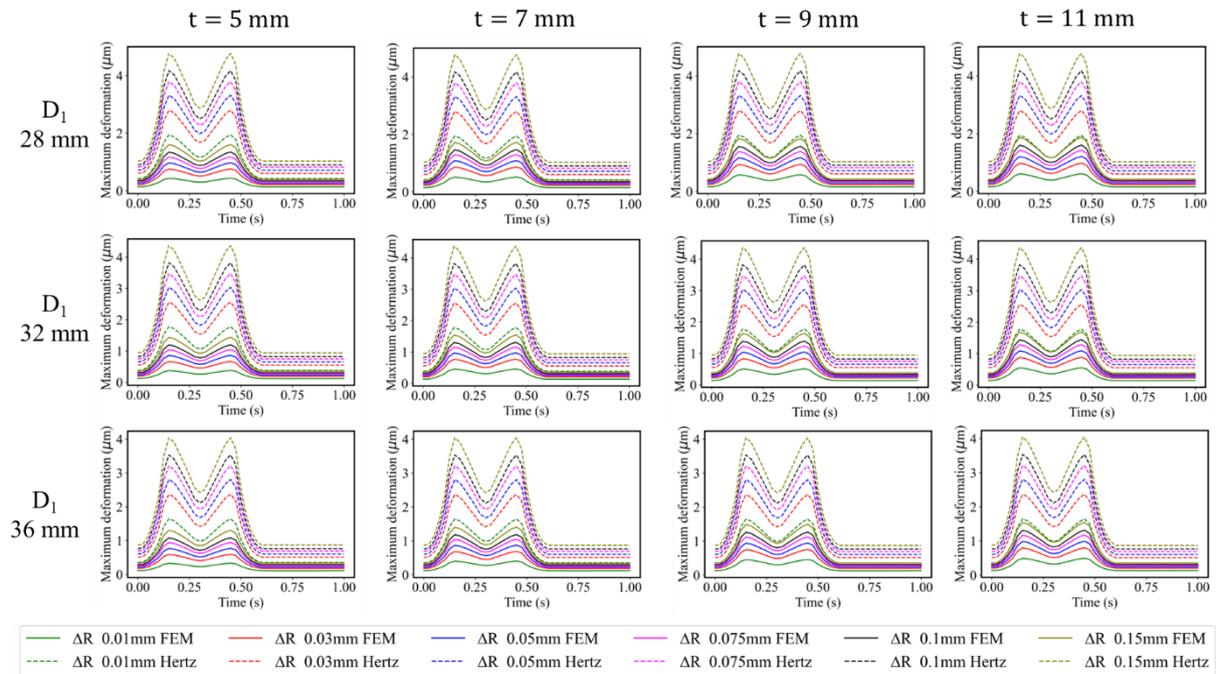

**Fig. C1.** Comparison of maximum elastic deformation over a single cycle between FEM and Hertz model for different radial clearances and thicknesses considering CoC tribo-pair

**Fang formulation**

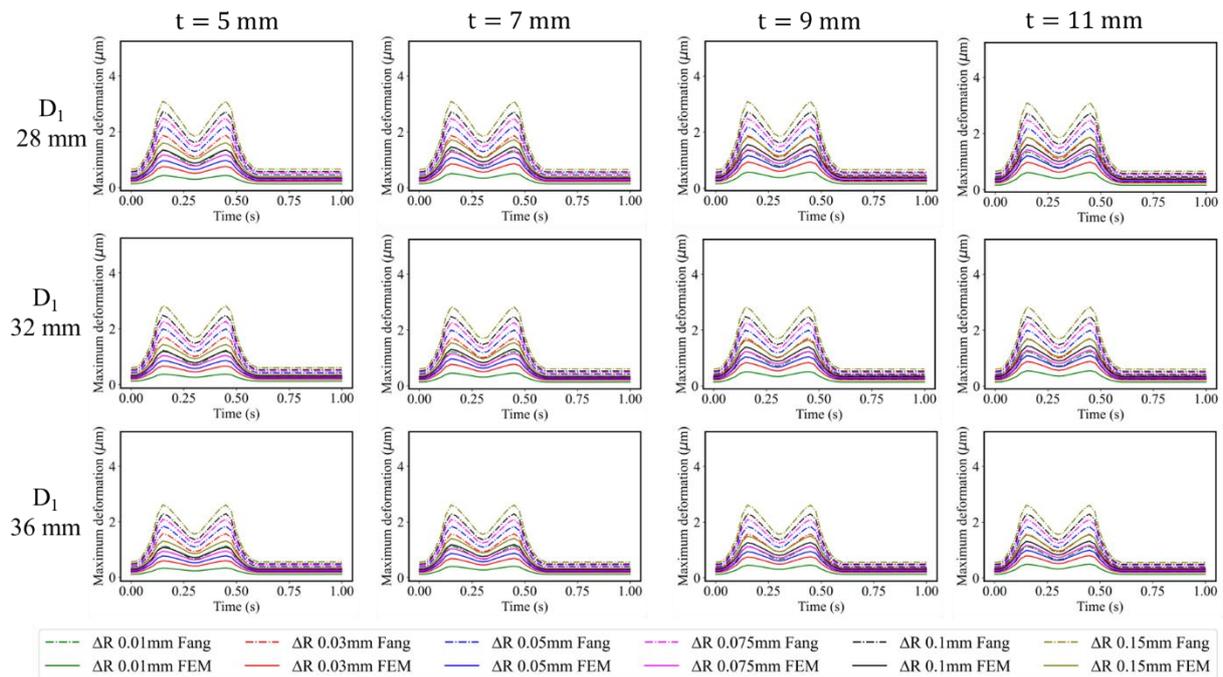

**Fig. C2.** Comparison of maximum elastic deformation over a single cycle between FEM and Fang model for different radial clearances and thicknesses considering CoC tribo-pair

## ORCID ID

K. Nitish Prasad: https://orcid.org/0000-0001-9906-4124

M. Abhilash: https://orcid.org/0009-0001-2148-8741

P. Ramkumar: http://orcid.org/0000-0002-2816-9145